\documentclass[12pt,usenatbib,iop]{emulateapj}
\newcommand{\calU}{{\cal U}}
\newcommand{\ppt}{{\partial\over\partial t} }
\newcommand{\ddr}{{\partial\over \partial r}}
\newcommand{\divg}{\nabla \cdot} 
\newcommand{\vect}[1]{\mathbf{#1}}
\newcommand{\frad}{{\mathbf f}_{\rm rad}}
\newcommand{\fmag}{{\mathbf f}_{\rm mag}}

\newcommand{\GammaiS}{\Gamma}
\newcommand{\rch}{\tilde{r}_{\rm ch} } 
\newcommand{\calUgal}{\left<{\cal U}_{\rm obs}\right>_{\rm gal}}
\newcommand{\ftrapw}{f_{{\rm trap},w}}   
\newcommand{\HII}{\ion{H}{2} }
\newcommand{\ArIII}{Ar\,{\footnotesize III}}
\newcommand{\ArII}{Ar\,{\footnotesize II}}
\newcommand{\SIV}{S\,{\footnotesize IV}}
\newcommand{\SIII}{S\,{\footnotesize III}}
\newcommand{\NeIII}{Ne\,{\footnotesize III}}
\newcommand{\NeII}{Ne\,{\footnotesize II}}

\begin{document}

\slugcomment{Accepted for publication in ApJ, June 26, 2012}

\title{Ionization Parameter as a Diagnostic of Radiation and Wind Pressures in \HII Regions and Starburst Galaxies} 
\author{Sherry C. C. Yeh \& Christopher D. Matzner}
\affil{Department of Astronomy \& Astrophysics, University of Toronto, 50 St. George St., Toronto, ON M5S 3H4, Canada}

\begin{abstract}
The ionization parameter $\calU$ is potentially useful as a tool to measure radiation pressure feedback from massive star clusters, as it directly reflects the ratio of radiation to gas pressure and is readily derived from mid-infrared line ratios.   We consider a number of physical effects which combine to determine the apparent value of $\calU$ in observations encompassing one or many \HII regions.   An upper limit is set by the compression of gas by radiation pressure, when this is important.  The pressure of shocked stellar winds and the presence of neutral clumps both tend to reduce $\calU$ for a given intesity of irradiation.     The most intensely irradiated regions are selectively dimmed by internal dust absorption of ionizing photons, leading to a bias for observations on galactic scales.  We explore these effects in analytical and numerical models for dusty \HII regions and use them to interpret previous observational results.  
We find that radiation pressure confinement sets the  upper limit $\log_{10} \calU \simeq-1$ seen in individual regions.  Unresolved starbursts are known to display a maximum value of $\simeq -2.3$.  While lower, this is also consistent with a large portion of their \HII regions being radiation pressure dominated, given the different technique used to interpret unresolved regions, and given the bias caused by dust absorption.  
We infer that many individual, strongly illuminated regions cannot be significantly overpressured by stellar winds, and that even when averaged on galactic scales, the shocked wind pressure cannot be large compared to radiation pressure.  Therefore, most \HII regions cannot be adiabatic wind bubbles.  Our models imply a metallicity dependence in the physical structure and dust attenuation of radiation-dominated regions, both of which should vary strongly across a critical metallicity of about one-twentieth solar. 
\end{abstract}
\keywords{\HII regions --- galaxies: starburst --- galaxies: general --- ISM: bubbles --- dust, extinction ---  infrared: galaxies} 

\section{Introduction}

Forbidden line ratios are often employed to determine the hardness of the radiation field and the composition and physical state of irradiated gas within star-forming galaxies. 
Of the quantities revealed by these ratios, the ionization parameter $\calU=n_{\gamma,i}/n_H$ -- the ratio of ionizing photon density to hydrogen density -- is of special interest, as it measures the dimensionless intensity of ionizing radiation.  Because $\calU$ controls the 
ionization state \citep[e.g.,][]{1969ApJ...156..943T}, it is easily disentangled from other parameters \citep{dopita00}; and because it indicates the ratio of ionizing radiation pressure to gas pressure \citep[e.g.,][]{1999agnc.book.....K}, it encodes information about the relative importance of these forces in star formation feedback.  

It is important, therefore, to understand the physical and observational effects which influence our estimates of $\calU$ in distant galaxies.  When  determined on kiloparsec scales, $\log_{10}\calU$ appears to reach a maximum value of about -2.3 in starburst environments, and indeed holds this value throughout the inner 500 pc of M82 \citep{thornley00,fs01, carral94,smith06, west07}.  These points led \citet{thornley00} to argue that this maximum $\log_{10}\calU$ is a property of the starburst phenomenon.   However, higher values (up to about -1) are sometimes seen in cases where \HII regions are individually resolved \citep{snijders07,indeb09}.     

Why do individual \HII regions not exhibit arbitrarily high ionization parameter, and why are starburst determinations limited to $\log_{10}\calU \leq {-2.3}$?  Both limits strongly suggest the operation of a saturation mechanism.  We know of four possibilities: 

{\em Dynamics.} The characteristic values of $\calU$ could originate in the dynamics of expanding \HII regions.  \citeauthor{dopita05}\ (\citeyear{dopita05}, hereafter D05), in particular, explain the saturation of $\calU$ in terms of the `stalling' of  expanding \HII regions as they decelerate to the ambient sound speed.  Under a specific set of assumptions about the dynamics of \HII regions and their interaction with ambient gas, this implies a unique value of $\calU$ at the time of stalling, which imprints itself on the line ratios integrated over an entire galaxy.  However, there are two problems with the assumptions that enter this explanation.  

First, D05 follow \citet{weaver77} in modeling \HII regions as though they were enclosed stellar wind bubbles and therefore use the \citeauthor{weaver77}\ model to describe their motions.   Two pieces of evidence indicate that wind energy escapes \HII regions,  contrary to this assumption.  Trapped wind would produce x-ray luminosities as much as two orders of magnitude higher than what is observed \citep[e.g., in the Carina Nebula:][]{hc09}.  Moreover, D05 find that the wind luminosity must be drastically scaled down to reproduce the observed range of $\calU$ values.  (We revisit this issue below in \S \ref{SS:Winds} and \S \ref{conclusion}.)

Second, D05 assume that the ambient gas is the warm (ionized or neutral)  component of the ISM, and therefore has a unique sound speed of roughly $10$\,km\,s$^{-1}$.  Although this is plausible, many of the ionizing photons will be consumed as the dense molecular material or cold neutral hydrogen is being cleared away.  As the thermal sound speed of this matter is quite low, the appropriate stalling criterion involves the total (turbulent, magnetic and thermal) ambient pressure \citep[e.g.,][]{1993ApJ...417..187S,matzner02}.

{\em Homogeneous mixtures.} A separate explanation has been provided by \citet[][FS01]{fs01}  in the context of the central 500\,pc starburst core of M82, where their and later observations \citep{smith06,west07} demonstrate that $\calU$ is remarkably uniform.  
The FS01 model invokes an ensemble of neutral clouds with ionized outer layers, scattered among newly formed stars.  The uniformity of $\calU$ must therefore be sought in the properties of this mixture, which are not part of the \citeauthor{fs01} model.   For this reason we do not investigate mixture models further, except to point out that most physical mechanisms limiting $\calU$ in individual \HII regions would also operate here. 

{\em Radiation pressure confinement}. Another physical mechanism for the saturation of $\calU$ was introduced by \citet{binette97} and elaborated by \citeauthor{dopita02}\ (\citeyear{dopita02}, D02) in the context of narrow-line regions in active galactic nuclei.   When irradiation is sufficiently intense, ionized gas develops a pressure gradient to oppose the force of photons caught by neutral atoms and dust grains, and is squeezed into a thin layer near the ionization front (IF).    This prevents gas pressure from becoming small compared to the ionizing radiation pressure (as previously argued by \citealt{ferland84}).   The corresponding upper limit on $\log_{10} \calU$ depends on the ionizing spectrum and the lines used to infer ionization; 
a rough estimate based on pressure equilibrium gives -1.5.   While this is significantly above the value of -2.3 seen in starbursts,  the details of line formation, among other effects, can reduce $\calU$ to about its observed maximum.  

{\em Dust-limited \HII regions}.  Finally, the observed saturation could simply be a selection effect, caused by the fact that dust absorbs most of the ionizing starlight in regions with $\log_{10} \calU\gtrsim -2$, suppressing their line emission.   Because extragalactic observations inevitably integrate over a wide range of local conditions, the inferred $\calU$ will be weighted toward regions of lower intensity.  Such a `dust-limited' state has been shown to successfully reproduce the suppression of fine structure lines relative to far-infrared luminosity in ultra-luminous infrared galaxies \citep{1992ApJ...399..495V,1998PASP..110.1040B,2009ApJ...701.1147A}.   
 
Our goal in this paper is to evaluate these possibilities and, in the process, to assess whether $\calU$ can be a useful proxy to measure radiation pressure feedback in \HII regions in external galaxies. 
We review the physical effects which control the state of ionized gas and the values of $\calU$ inferred from forbidden line ratios, focusing on its saturation in starburst galaxies and the relevance of the last two mechanisms.  Because radiation pressure and dust opacity are both proportional to $\calU$, they occur simultaneously.  But, since they are truly two separate physical mechanisms, we take care not to conflate them.    In this paper we concentrate on the internal structures of \HII regions,  only evaluating a very simple model for their dynamics and galactic populations (\S \ref{SS:Dust_abs}).  A companion paper (Verdolini et al.\ 2012, in prep.) presents a new suite of models for the emission from galactic populations of \HII regions evolved with more sophisticated internal dynamics.

We begin in \S~\ref{S:Parameters} by reviewing the parameters of quasi-static, dusty, spherical \HII regions.  We draw here on the models of \citeauthor{draine11} (\citeyear{draine11}, hereafter Dr11), examining in detail their planar, radiation-confined limit (Appendix \ref{Appendix:PlanarRegions}), but we also introduce a new parameter to account for an inner, pressurized region of shocked stellar wind.   We discuss the scales of dust opacity and radiation pressure in \S~\ref{SS:Basics},  confinement by radiation pressure and the maximum $\calU$ in \S~\ref{SS:RadConfinement}, and the influence of an inner bounding pressure in \S~\ref{SS:PressureParameter}. 

In \S~\ref{S:Dynamics} we focus on the extra effects which can reduce the maximum value of $\calU$, including 
stellar winds and neutral clumps in the region. Magnetic fields can in principle allow $\calU$ to violate its nonmagnetic upper limit, but this turns out to be difficult to arrange. We discuss the effects of internal and external dust extinctions on limiting $\calU$ in \S~\ref{SS:Dust_abs}.
In \S~\ref{S:model}, we provide a survey of $\calU$ values obtained from modeled MIR line ratios within static and internally pressurized
dusty \HII regions, and demonstrate how stellar winds act to suppress $\calU_{\rm obs}$.
In \S~\ref{S:implication}, we use our results to interpret observations on the scales of individual \HII regions and entire galaxies.

\section{Quasi-Static \HII regions: Parameters, Properties, and Saturation of $ {\calU}_{\rm obs}$} \label{S:Parameters} 

In the static or quasi-static state, \HII regions have several well-known limits: classical \HII regions \citep{stromgren39}, which lack significant 
radiation pressure or dust opacity; dusty \HII regions \citep{mathis71,petrosian72} in which radiation pressure is negligible; 
and radiation-confined \HII regions \citep{binette97,dopita02} in which both radiation pressure and dust opacity are generally important.    
Recently, Dr11 has thoroughly explored the parameter space of dusty \HII regions with radiation pressure, 
demonstrating a three-parameter family of models which takes all of these limits.  

Ionization and force balance within spherical \HII regions are especially simple under the following set of 
idealizations (Dr11): 
the ionized gas temperature $T$, the mean energy of ionizing photons $\bar e_i = \left<h\nu\right>_i$, the effective dust 
opacity per H atom $\sigma_d$, and the effective recombination coefficient $\alpha$, are all uniform constants; 
pressure from reprocessed radiation, such as  
the Lyman $\alpha$ line, is unimportant; 
the flow  is static, so forces must balance; 
the pressure from stellar winds is neglected (which we will relax below in \S \ref{SS:PressureParameter} and \S \ref{SS:Winds}); and 
there are no additional forces from gravity or magnetic fields.   

With these idealizations, \HII regions are described by seven dimensional parameters: 
 $S$, the central output of ionizing photons per second; 
 $L_i/c$ and $L_n/c$, the total force of ionizing and non-ionizing photons\footnote{$L_n$ refers to the frequency range where the dust opacity is still $\simeq \sigma_d$;  Dr11 assumes that the two opacities are equal.};
$\alpha$, the effective recombination coefficient;
$P/n_H = 2.2 k_B T$, the thermal pressure\footnote{Ignoring helium, Dr11 takes this coefficient to be 2; our value of 2.2 assumes He is singly-ionized everywhere.} per H density;
$P_{\rm IF}$, the pressure at the ionization front; and  $\sigma_d$.

Other dimensional parameters of interest, determined by the seven above, include: $n_{i*}= P_{\rm IF}/(2.2 k_B T)$, the H density at the ionization front; $R_{\rm St} = [3S/(4 \pi \alpha n_{i*})]^{1/3}$, the radius of a dust-free Str\"omgren sphere of uniform density $n_{i*}$;  $n_{\rm em}$, the recombination-weighted ion density (eq.~[\ref{def:n_em}]),  and 
\begin{equation} \label{eq:Rch} 
\rch = {\alpha L^2 \over 12\pi  (2.2 k T c)^2 S}, 
\end{equation}
the radius of a uniform-density, dust-free Str\"omgren sphere whose gas pressure equals the total, unattenuated radiation pressure $L/(4\pi R_{\rm St}^2 c)$ at its boundary.   (Our $\rch$ differs from \citealt{km09}'s [hereafter KM09's] $r_{\rm ch}$ in that it accounts neither for the effects of dust extinction, nor for the additional pressure from stellar winds.)  Yet another is  $R_{\rm IF}$, the ionization front radius. 

We focus on three dimensionless ratios which control the structure and appearance of the \HII region.\footnote{A fourth, the total number of H atoms in the classical Str\"omgren sphere, is unimportant.}  First of these is the characteristic ratio of ionizing radiation pressure to gas pressure in a classical Str\"omgren sphere,
\begin{eqnarray}  \label{eq:GammaDefn}
\GammaiS &\equiv& {S \bar e_i \over 4\pi R_{\rm St}^2 c (2.2 n_{i*} k_B T)} = {L_i \over L} \left(\rch\over R_{\rm St} \right)^{1/2}\\ 
&~&= {0.93\over T_4^{1.53}} (S_{51} n_{i*,4})^{1/3} , \nonumber
\end{eqnarray}
where n$_{i*}$=10$^4$ n$_{i*,4}$ cm$^{-3}$, $S=10^{51} S_{51}$\,s$^{-1}$, and we have used $\alpha = 2.54\times10^{-13} T_4^{-0.81}$\,cm${^3}$\,s$^{-1}$ \citep{1995MNRAS.272...41S}.   Here and elsewhere, we use the subscript `St' to denote conditions in a classical, uniform Str\"omgren sphere without attenuation, rather than the actual \HII region.   

An alternative to $\GammaiS$ foregoes $R_{\rm St}$ as a dimensional scale in favor of the actual ionization front radius $R_{\rm IF}$; 
\begin{equation}\label{Psi_definition} 
\Psi \equiv {R_{\rm IF}\over \rch}.  
\end{equation} 
This is useful if $R_{\rm IF}$ is observed or derived from the dynamics of the region (KM09); we use $\Psi$ to parameterize our simulation output in \S~\ref{S:model}.   Bear in mind that $\Psi$ decreases as radiation pressure becomes more important, whereas $\GammaiS$ increases. 

Our second parameter is the dust opacity in a classical \HII region,
\begin{equation} \label{eq:taudDefn}
\tau_{d,{\rm St}} = \sigma_d n_{i*} R_{\rm St}. 
\end{equation}
Both $\GammaiS$ and $\tau_{d,{\rm St}}$ are proportional to $(Sn_{i*})^{1/3}$; as a result $\tau_d\propto \calU$ in uniform \HII regions. 

The third parameter is the {\em dust discriminant}
\begin{eqnarray} \label{eq:gamma_in_Dr11} 
\gamma &\equiv&  {2.2 k_b T  \over \bar e_i} {\sigma_d c\over \alpha}  =   {\tau_{d,{\rm St } }\over3 \GammaiS}\\ 
&=& 16.3 \sigma_{d,-21} T_4^{1.8} {13.6\,{\rm eV}\over \bar e_i} \nonumber 
\end{eqnarray} 
where $\sigma_d = 10^{-21} \sigma_{d,-21}$\,cm$^{2}$.   This sets the relative importance of dust and neutral H atoms in absorbing ionizing photons when radiation pressure confines the gas: in a situation where gas pressure matches the ionizing radiation pressure, the dust absorbs a fraction $\gamma/(1+\gamma)$ of all ionizing photons.   Fiducial parameters ($\sigma_{d,-21}=1, T_4=0.8, \bar e_i = 20$\,eV) give $\gamma=7.4$.  

Additional parameters are required to capture the spectral form of the incident starlight and dust opacity, the most relevant of which,  $\beta$, compares the radiation force of non-ionizing starlight to that of ionizing starlight; see Dr11 and Appendix \ref{Appendix:PlanarRegions}. 

For emission in a given line, and therefore for the determination of line ratios and electron temperatures, the ratio of the characteristic density $n_{i*}$ to the transition's critical density $n_{\rm crit}$ is a significant additional parameter.  So long as this ratio is small for all of the lines which are either observed or contribute to the collisional cooling of the region, the density parameter will not be important.   We return to this point in \S \ref{S:model}. 

Although the above idealization of \HII regions is reasonable
and admits a simple parameter space, we wish to relax its approximations and assumptions when considering observational constraints. 
Microphysical assumptions, such as the constancy of $T$, can be relaxed through the use of a modern photoionization code like Cloudy or MAPPINGS; 
we take this approach in \S~\ref{S:model}.   Macrophysical issues, such as 
stellar winds and bulk flow, must be addressed separately (\S~\ref{S:Dynamics}).   

\subsection{Ionization parameter, radiation pressure, and dust opacity} \label{SS:Basics} 

We are especially interested in observational measures of the ionization parameter ${\calU}({\mathbf r}) = n_{\gamma i}/n_H$.
This is directly related to  the ionizing radiation-to-gas pressure ratio 
\begin{eqnarray} \label{eq:Xi}
\Xi_i({\mathbf r}) &\equiv& \frac{P_{{\rm rad},i}({\mathbf r})}{P_{\rm gas}({\mathbf r})} = 
\calU({\mathbf r}) \frac{\bar e_i}{2.2 k_B T}. 
\end{eqnarray}

These are both functions of position $\mathbf r$, and are each observable only in an integral sense within the region.   Whereas ${\calU}(\mathbf r)
$ and the ionizing spectrum set the local ionization structure and recombination spectrum, any recombination line ratio will depend on an average 
of $\calU$ weighted by the emissivities of the species involved. The value ${\calU}_{\rm obs}$ inferred from such line ratios will naturally 
reflect this weighting.   One observationally relevant proxy is therefore the recombination-weighted average
\begin{equation} \label{eq:U_em} 
{\calU}_{\rm em} = {\int n_H^2 \, {\calU}\, dV \over \int n_H^2\, dV}, 
\end{equation} 
where both integrals are over the entire \HII region.   Another commonly-used proxy is 
a `geometrical' estimate which relies on an observational determination of the characteristic ionized hydrogen density: 
\begin{equation} \label{Ugeom} 
\calU_{\rm geom} = {S\over 4\pi R_{\rm IF}^2 n_{i,{\rm obs}} c}.
\end{equation} 
We also define $[\Xi_{i,{\rm em}}, \Xi_{i, {\rm geom}}] $= $ (\bar e_i/2.2k_BT)[{\calU}_{\rm em}, \calU_{\rm geom}]$ for consistency. In a classical \HII region (uniform density, negligible dust absorption, radius $R_{\rm St}$),
\begin{equation}\label{eq:Xiem_vs_Gamma_in_Classical} 
 \Xi_{i,{\rm em}}  = {9\over 4} \GammaiS. 
 \end{equation} 
and, in such a region, $\Xi_{i,{\rm geom}} = \GammaiS$ by definition. 

\begin{deluxetable*}{clcl}  \tabletypesize{\scriptsize}
\tablewidth{0pt}
\tablecaption{Definitions
}
\tablehead{
\colhead{Symbol} & 
\colhead{Definition} &
\colhead{Reference} &
\colhead{Note}}
\startdata
$[\Xi, \Xi_i]({\mathbf r})$ & Ratio of [total, ionizing] radiation pressure to gas pressure & Eq.\ (\ref{eq:Xi}) & Position dependent\\ 
$\calU({\mathbf r})$ & Ionization parameter $n_{\gamma,i}/n $& \nodata & Position dependent \\ 
$\calU_{\rm geom}$ & Geometric estimate $S/(4\pi R_{\rm IF}^2 n_{\rm obs} c)$ & Eq.\ (\ref{Ugeom}) & \\ 
$\calU_{\rm obs}$ & Observational estimate of $\calU$ for individual H\,II region or line of sight & \nodata & Tracer dependent\\ 
$\calU_{\rm obs,max}$ & Maximum value of $\calU_{\rm obs}$ in a quasi-static, unmagnetized layer &\nodata & Tracer dependnet \\ 
$\left<\calU_{\rm obs}\right>_{\rm gal}$ & Obs.\ estimate of $\calU$ for beams spanning many H\,II regions & Eq.\ (\ref{eq:Uobs_of_population}) & Biased by dust\\ 
$\calU_{\rm em}$ & Recombination-weighted average of $\calU$ & Eq.\ (\ref{eq:U_em}) & Theoretical proxy for $\calU_{\rm obs}$ \\
$\rch$ & Char. radius for radiation-gas pressure equality  & Eq.\ (\ref{eq:Rch}) & Similar to $r_{\rm ch}$ in KM09\\
$\Psi$ & Radiation pressure parameter, $R_{\rm IF}/\rch$ & Eq.\  (\ref{Psi_definition}) &$\Psi\gg1$ implies $P_{\rm rad}\ll P$ \\ 
 $\GammaiS$ &  Ionizing radiation pressure parameter, $\Psi^{-1/2} L_i/L$  & Eq.\ (\ref{eq:GammaDefn}) & $\GammaiS\ll1$ implies $P_{\rm rad}\ll P$\\ 
$\Omega$ & Wind pressure parameter, $P_{\rm in}V_{\rm in} / (P_{\rm IF} V_{\rm IF} - P_{\rm in} V_{\rm in})$& Eq.\ (\ref{define_Omega}) & $\Omega \gg1$ implies  $P_{\rm rad}\ll P$\\ 
$f_{{\rm trap}, w}$ & Wind pressure parameter, $4\pi R_{\rm in}^2 P_{\rm in} c/L$ & KM09 & \\ 
$f_w$ &Wind force parameter $ \dot{M}_w v_w c/L$  & \nodata& \\
$f_{\rm ion}$ & Fraction of ionizing photons absorbed by gas rather than dust & Dr11 & \\ 
$\tau_d$  & Dust optical depth to ion.\ photons from source to ionization front & Eq.\ (\ref{eq:taudDefn}) & $\tau_d > 1$: dust-limited H\,II region \\ 
$\gamma$ & Dust discriminant, $\propto \sigma_d T/(\bar e_i \alpha) $  & Eq.\ (\ref{eq:gamma_in_Dr11}), Dr11 & $\gamma > 1$ implies
              $\tau_d>1$ when rad-confined\\ 
$\beta$ & Effective ratio of non-ionizing to ionizing radiation force & Dr11 & See note in Table \ref{Table:SpectraParameters}\\
$n_{\rm em}$ & Recombination-weighted average density & Eq.\ (\ref{def:n_em}) & Proxy for observed density 
\enddata
\label{Definitions}
\end{deluxetable*}

These emission-weighted averages are directly related to the dust opacity, as the following argument shows.  
Following Dr11, we define the local 
ionizing photon density to be $S\phi(r)/(4\pi r^2 c)$, and the integrated recombination rate to be $\int \alpha n^2 \,dV = f_{\rm ion} S$, where $f_{\rm 
ion} = 1-f_{\rm dust}$ is the fraction of ionizing radiation absorbed by hydrogen rather than dust.   Rewriting both the numerator and denominator 
of equation (\ref{eq:U_em}), we find 
\begin{equation}\label{Xi_em-vs-tau_d}  
\Xi_{i,{\rm em}} = {\calU}_{\rm em} {\bar e_i\over 2.2 k_BT} ={\phi_d \over \gamma f_{\rm ion} } \tau_d
\end{equation} 
where 
\begin{equation}\label{<phi>_d} 
 \phi_d = {\int \phi(r) \sigma_d n_H dr \over \int \sigma_d n_H dr} 
\end{equation} 
is a dust absorption-weighted average of $\phi(r)$.     

Equation (\ref{Xi_em-vs-tau_d}) is remarkable, because it relates a region's dust optical depth to its ionizing photon-to-gas pressure ratio in terms of  
the dust discriminant $\gamma$, and the order-unity parameters $f_{\rm ion}$ and $\phi$ take definite forms in each of the limiting 
regimes of static \HII regions.    In the context of radiation-confined \HII regions, which we now discuss, $\gamma$ determines the 
importance of dust absorption. 

\subsection{Radiation confinement and saturation of   $\calU_{\rm obs}$} \label{SS:RadConfinement} 

In a classical \HII region ($\GammaiS\ll1, \Psi\gg1$), except for a small region near the ionizing source \citep{mathews67}, radiation pressure is negligible and has no practical effect on the distribution of ionized gas.  In this case the recombination-weighted radiation pressure parameter $\Xi_{\rm em}=(1+\beta) \Xi_{i,{\rm em}} $ is much less than unity, as can be seen from equation (\ref{eq:Xiem_vs_Gamma_in_Classical}).   

This ceases to be true as the illumination becomes more intense.    All of the momentum in ionizing starlight, and (thanks to dust) some of the momentum in non-ionizing starlight, is transferred to the ionized gas.  In a quasi-static region without other forces, the radiation force must be balanced by a gas pressure gradient.  Many have pointed out that when radiation pressure is significant ($\GammaiS \gg 1, \Psi\ll1)$, ionized gas can be effectively excluded from the interior  \citep{elitzur86} and confined to a thin shell near the IF \citep{binette97,dopita02,draine11}.  Within this shell the total pressure is nearly constant, as we show in Appendix \ref{Appendix:ForceBalance}, so the rise in gas pressure matches the drop in radiation pressure.   Although the ratio of radiation and gas pressures varies throughout the region, its characteristic value records a state of pressure balance  -- either $\Xi_{i, {\rm em} }\simeq 1$, if dust absorption is negligible ($\gamma<1$), or $\Xi_{\rm em} \simeq 1$ if it is not ($\gamma>1$).  We explore the radiation-confined limit more quantitatively in Appendix \ref{Appendix:PlanarRegions}. 

Because $\Xi_{\rm em}$ reaches its maximum value of around unity in radiation-confined layers, the recombination-weighted ionization parameter $\calU_{\rm em}$ does as well (Dr11): in a dusty layer with $\Xi_{\rm em} \simeq 1$, we expect $\calU_{\rm em} = 2.2 k_B T \Xi_{i,{\rm em}}/[(1+\beta)\bar e_i] \simeq 10^{-1.5}$.  (This estimate rises to $10^{-1}$ in low-dust regions with $\gamma<1$, because of the factor of $\Xi/\Xi_i = 1+\beta$.)    Moreover, the structure of the ionized zone becomes nearly independent of the strength of the illumination, apart from an overall scale, so long as $\GammaiS\gg1$. 

Any ratio of forbidden lines, which are all well above or well below their critical densities, therefore saturates at a finite value in this radiation-confined limit \citep[e.g.][]{binette97}.   The ionization parameter $\calU_{\rm obs}$ which would be inferred from a specific set of line ratios tends to be close, but not equal, to $\calU_{\rm em}$.  Because the radiation force causes $\calU(r)$ to be strongly stratified, there exists a diffuse inner region where lines preferentially form that require a higher ionization state \citep[][]{dopita02}. Conversely, low-ionization transitions form in outer regions of lower $\calU$.  The observationally-inferred value ${\calU}_{\rm obs}$ therefore limits to a maximum value which depends on the lines being observed in addition to all the parameters apart from the strength of illumination:  the chemical abundances, dust properties, and incident spectrum.   In other words 
\begin{equation} 
{\calU}_{\rm obs} \leq {\calU}_{\rm obs, max}, 
\end{equation} 
where $\calU_{\rm obs, max}$ depends on the lines in question and these other parameters, but not on the radiation intensity (unless radiation pressure leads the density to cross one of its critical values).   Likewise, if density-dependent tracers are used to derive $n_{i,{\rm obs}}$ and $\calU_{\rm geom}$, then this will exhibit an upper limit in the vicinity of $\max(\calU_{\rm em})\simeq 10^{-1.5}$ in the radiation-confined state.  In  \S~\ref{S:model} we demonstrate the saturation of $\calU_{\rm obs}$ for a selection of mid-infrared emission lines with high critical densities using the Cloudy code.  We find in Figure \ref{fig:ne3ne2_2myr} that $\calU_{\rm obs}$ derived from a particular [\NeIII]/[\NeII] line ratio exhibits a maximum at $10^{-1.97}$ for our fiducial parameters. 

Because of the effects just described, the saturation of emission line ratios is a feature of the radiation-confined state in which photon momentum dominates over gas pressure gradients in the structure and dynamics of \HII regions.   As we shall see in \S \ref{SS:Dust_abs}, however, a selection effect due to dust attenuation (\S~\ref{SS:Dust_abs}) also modulates the apparent value of $\calU_{\rm obs}$ in large-scale observations of dusty galaxies ($\gamma>1$).  Both effects are important in the interpretation of galactic line ratios. 

In Appendix \ref{Appendix:PlanarRegions} and Figures \ref{TaudXiFion} and \ref{XiEstimatesPlanar}, 
we present the properties of \HII regions in the radiation-confined state, i.e., the planar limit of Dr11's equations in the absence of a finite pressure at the inner boundary. This limit, which corresponds to $\GammaiS\gg1$ (and $\Omega\ll1$ in the definitions of \S \ref{SS:PressureParameter}), is parameterized only by the dust discriminant $\gamma$ and the spectral parameter $\beta$, which quantifies the non-ionizing radiation pressure. Although these solutions are only a subset of Dr11's three-parameter model family, they improve upon the treatment by \citet{dopita03} by accounting for the gas density gradient induced by radiation pressure.  They show that: (1) $\Xi_{i,{\rm em}}$ saturates at a value which is near unity and only logarithmically dependent on $\gamma$.  (2) Dust absorbs most ionizing radiation, so  $\tau_d\sim 1-3$ for astrophysically relevant values of $\gamma$.    
As Dr11 discusses, $\gamma \sim 10$ at solar metallicity unless dust is effectively removed. Therefore radiation-confined \HII regions are also 
marginally `dust-limited'; see \S\ref{SSS:Int_abs}.  (3) For relatively dust-free gas with $\gamma<1$, the planar solution is not compact in radius.  It must either be bounded inward by a bubble of hot gas, or be matched to a spherical solution (as in Dr11).\footnote{Along with the relatively short dust drift time scale (Dr11), this suggests that dust-free gas may escape inward of the dusty layer even when $\gamma$ is initially large.  We do not account for this effect in our equations and models, as we assume a uniform mixture of gas and dust.} 

In the radiation-confined shell, the column density of ionized gas takes a characteristic maximum value: either $\simeq 2.2 k_BTc/(\alpha \bar e_i)$ if dust absorption is negligible, or several times $\sigma_d^{-1}$ if it is not.  What happens if the mean column density of the ambient medium is less than four times this shell column, so there is not enough matter to achieve it?  In this case, it is easy to show that the initial Str\"omgren sphere forms outside $\tilde r_{\rm ch}$, so that $\Psi>1$ and $\GammaiS<1$: the radiation-dominated phase never occurs.

\subsection{Pressurized inner boundary}  \label{SS:PressureParameter}
The ionizing sources of \HII regions (typically stars, star clusters, or AGN) often produce fast outflows which decelerate or are deflected before they reach the IF, thereby creating regions of hot and pressurized gas within the photo-ionized zone.  What new parameter should we introduce to capture the effect of such zones on the \HII regions' emission? 
The ratio of outer to inner pressures is not appropriate, because the two are equal in a classical \HII region and also in a wind-confined shell.   Neither is the volume fraction of the interior zone, because photo-ionized gas can be confined to a thin shell by radiation pressure just as well.  Instead we choose 
\begin{equation}\label{define_Omega}
\Omega \equiv \frac{P_{\rm in}V_{\rm in}}{P_{\rm IF}V_{\rm IF} - P_{\rm in}V_{\rm in}}~~,
\end{equation}
where $
P_{\rm IF}V_{\rm IF} - P_{\rm in}V_{\rm in}$ is the difference of the product of pressure and volume between the IF and the inner edge of the \HII region (the outer edge of a pressurized bubble).    This combination reflects the strength of the interior pressure, both for classical and radiation-dominated \HII regions.  In fact $\Omega$ directly reflects the contribution of the interior bubble to the total kinetic energy of the region, which, according to the virial theorem \citep{mckee92}, controls the expansion of it and any shell of neutral matter it pushes outward (e.g., KM09).   

\begin{figure}
\epsscale{1}
\plotone{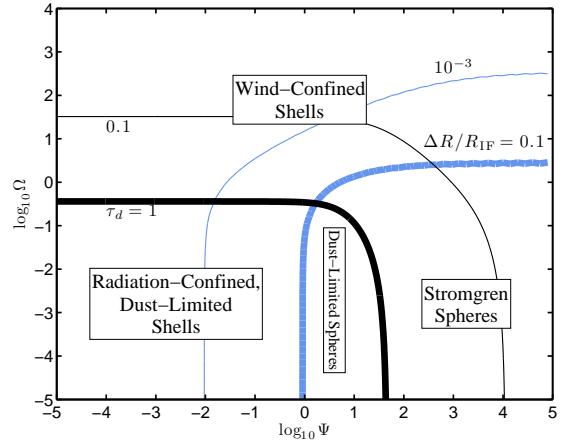}
\caption{ Parameter space of quasi-static, spherical, unmagnetized, dusty \HII regions surrounding central wind bubbles.  Wind-confined, radiation-confined, dust-limited, and classical (Str\"omgren) regions are indicated.  Black contours denote the radial dust optical depth $\tau_d$ from the source to the ionization front; blue contours indicate the radial thickness of the ionized zone, $\Delta R/R_{\rm IF}$.   We use thick contours to separate fiducial regimes. 
For this figure we adopt $\beta = 1$, $\gamma = 7.4$; the radiation-confined state is encompassed by the dust-limited state so long as $\gamma > 1$.   }
\label{fig:ParameterSpace}
\end{figure}

In Figure \ref{fig:ParameterSpace} we indicate the physical regimes of spherical, quasi-static, dusty \HII regions surrounding a central wind bubble. 

Quasi-static HII regions typically have an inward gradient of $\calU$. For classical \HII regions this is simply because the ionizing flux is greater at smaller radii, while for radiation-confined regions it is due to the gas density gradient.  In either case, replacing a region's inner portion with hot gas has the effect of removing its most highly photo-ionized zone.   This effect will be greatest for those lines more which form preferentially in regions of highest $\calU$. See \S \ref{SS:Winds} for a discussion of the effect on $\calU_{\rm obs}$.

In radiation-dominated ($\GammaiS\gg1$) and relatively dust free ($\gamma \leq 1$) regions,  low-density ionized gas fills the interior (Appendix \ref{Appendix:PlanarRegions}). Because this low-density gas is ideal for the formation of lines sensitive to regions of high $\calU$, we expect these line ratios to be especially sensitive to interior pressure when $\GammaiS\gg1$ and $\gamma\leq1$.   

\section{Factors affecting ${\calU}_{\rm obs}$ in individual \HII regions}\label{S:Dynamics} 

Several physical effects modify ${\calU}_{\rm obs}$ relative to its  theoretical maximum value $\calU_{\rm obs,max}$ within an individual \HII region. In the following subsections we consider stellar winds, neutral clumps, dust absorption of non-ionizing light, and magnetic fields.  Of these, only magnetic fields act to raise  $\calU_{\rm obs}$  relative to $\calU_{\rm obs,max}$.  Although highly-magnetized regions can in principle have  ${\calU}_{\rm obs}> \calU_{\rm obs,max}$, we shall see that this is very difficult to arrange in practice.  

\subsection{ Pressurization by shocked stellar winds} \label{SS:Winds} 

As discussed above in \S~\ref{SS:PressureParameter}, any force which clears ionized gas from the interior of an \HII region will have the effect of reducing the characteristic value of ${\calU}_{\rm obs}$, especially for those lines which form in regions of highest $\calU$.  The most relevant force is the pressure of a shocked wind produced by the same source that ionizes the region.  Most such regions are ionized by hot stars, which inject fast line-driven winds.   However the significance of stellar wind pressure depends on the wind strength, e.g. the ratio $f_w = \dot{M}_w v_w c/L$ between the momentum fluxes of winds and starlight, and the trapping of wind energy within the region.   KM09 combine these into a single parameter $f_{{\rm trap},w}\equiv 4 \pi P_{\rm in} R_{\rm in}^2 c/L \geq f_w$ which encapsulates the contribution from shocked winds relative to the direct momentum flux of starlight.   

So long as the combined force of winds and radiation compresses the ionized layer into a thin shell, the enclosed volume is constant and $\Omega$ reduces a comparison of pressures. Since $P_{\rm IF} \simeq P_{\rm in} + L/(4\pi r^2 c)$ in this case (Appendix \ref{Appendix:ForceBalance}, ignoring on the last term the factor $[1+\beta(1-e^{-\tau_d})]/(1+\beta)$ which accounts for incomplete absorption), we expect 
\begin{equation}\label{OmegaFromFtrapw}
\Omega \simeq 4\pi r^2 c P_{\rm in}/L = f_{\rm trap,w}.
\end{equation}
Further, if we consider a typical location where half of the incident radiation has been absorbed and its pressure applied to the gas, the same approximations lead to the estimate 
\begin{equation} \label{XiFromFtrapw}
\Xi \simeq {1\over 2f_{{\rm trap},w}+1},
\end{equation}    
although, in the case where confinement is accomplished by radiation alone ($f_{{\rm trap},w}\ll1$), the values of $\Xi_{\rm em}$ plotted in Figure \ref{XiEstimatesPlanar} are more accurate. 

We refer to section 3.1 of KM09 for a detailed discussion of the characteristic values of $f_w$ and $\ftrapw$ one might expect around massive star clusters.   For our purposes, it suffices to consider three limiting cases: (1) perfectly trapped, adiabatic bubbles of shocked wind, as envisioned in the models of \citet{weaver77} and \citet{km92b} and employed by \citet{dopita05,dopita06model}, which have $\ftrapw \sim f_w v_w/\dot{R}_{\rm IF} \gg 1$; (2) leaky wind bubbles (\citealt{hc09}, KM09), for which  $\ftrapw\sim 1$ in the case of a very massive cluster of solar metallicity (KM09); and (3) very weak wind bubbles with $\ftrapw\ll 1$, such as one might expect if the central source consists of low-metallicity or lower-mass main sequence stars.  

In the first case (adiabatic wind bubbles) $\Omega\gg1$ and  ${\calU}_{\rm obs}$ will be limited to values well below ${\calU}_{\rm obs, max}$.   Indeed, $\calU_{\rm geom} $ can easily be computed from the self-similar solutions of \citet{km92b}: using $L = (1+\beta) S \bar e_i$ and evaluating for the case of a steady wind in a uniform medium \citep{weaver77}, 
\begin{equation} \label{UgeomInWeaver77} 
\calU_{\rm geom} = 2 \calU_{\rm em} = {2.2 k_B T_i \over 3 (1+\beta)\bar e_i } {\dot{R}_{\rm IF} \over f_w v_w}.
\end{equation} 
The factor of two in the first equality arises because a pressure-confined ionized layer is nearly uniform and optically thin to dust; therefore $\calU$ drops linearly in radius from the value $\calU_{\rm geom}$ at its inner edge. 
Since $v_w\simeq 10^3$\,km\,s$^{-1}$ from a young cluster which fully samples the IMF, equation (\ref{UgeomInWeaver77}) yields $\calU_{\rm geom} \simeq 10^{-3.8}f_w^{-1}  (\dot R_{\rm IF}/10\,{\rm km\,s^{-1}}) $.  Equation (\ref{UgeomInWeaver77}) assumes the ionization front is trapped within the swept-up ambient matter, which requires  
\begin{equation}
S_{49} n_{H0} >  1.5 {T_{4}^{3.6}\dot{R}_{\rm IF6}^5  \over (1+\beta)^3 f_w^3 v_{w8}^3} \left(20\,{\rm eV}\over \bar e_i\right)^3
\end{equation}
where $n_{H0}$ is the ambient H density in cm$^{-3}$, $\dot{R}_{\rm IF} = 10\dot{R}_{\rm IF6}$\,km\,s$^{-1}$, and $v_w = 10^3 v_{w8}$\,km\,s$^{-1}$. 

In the second case (leaky bubbles), the wind force is of order $\ftrapw L/c$ and is therefore significant when $\ftrapw$ exceeds unity.  So long as the ionized layer is thin, the suppression of $\calU_{\rm obs}$ relative to $\calU_{\rm obs,max}$ is of order $\Xi$ as given by equation (\ref{XiFromFtrapw}).

 In the last case (very weak winds), $\calU_{\rm obs}$ will be unaffected by their presence, although high-$\calU$ lines from the innermost regions may be (especially when $\gamma\leq 1$).    In our modeling effort (\S~\ref{S:model}), we consider a wide range of possible values for $\Omega$. 

Spherically symmetric, partially-radiative bubbles \citep{km92b,mk08}, which have $\ftrapw$ lower than the adiabatic case but well above unity, are not especially relevant here because they require slower winds than is typical of ionizing stars.    Fast-wind bubbles can also become radiative, however, because of mass entrainment from evaporating clumps (see \S \ref{SS:Clumps}).   Note that \citet{mbl84} argued that this effect prevents the filling factor of shocked wind, and therefore also $\ftrapw$, from exceeding unity. 

As an aside, we note that some ultracompact \HII regions are probably formed in quasi-static layers confined by the ram pressure of protostellar winds, which are both highly collimated and much stronger than radiation pressure.  Combining the outflow force-luminosity correlation reported by \citet{wu04} with the collimation model of \citet{matzner99}, 
\begin{equation}\label{fw_in_protostellar_jets}
f_w \simeq 7.4 {(10^4 L_\odot/L)^{0.35} \over \sin(\theta)^2 + 10^{-4} }
\end{equation} 
for an angle $\theta$ from the wind axis and a bolometric source luminosity $L$.  Such regions are typically in the first, wind-dominated regime, even well away from the wind axis.   This ram-pressure confined phase can only exist in directions where the wind itself is fully ionized.  The alternate case, in which the IF forms in the wind, was considered by \citet{2003IAUS..221P.274T}.  It is inertially-confined in the sense of \S \ref{InertialConfinement} below, and can exhibit $\calU_{\rm obs} > \calU_{\rm obs, max}$. 

\subsection{Clumps within \HII regions} \label{SS:Clumps} 
\HII regions are often porous and clumpy instead of spherical, as we invoked above in the context of wind trapping.  Here we wish to point out that clumpiness reduces the inferred $\calU_{\rm obs}$ relative to a spherical region of the same radius.  The net effect is therefore to reduce $\calU_{\rm obs}$ relative to $\calU_{\rm obs, max}$, although the upper limit does not necessarily change.  

In the absence of significant radiation pressure, a clump subjected to ionizing radiation is first crushed by the recoil of its photo-evaporative flow 
\citep{bertoldi89}. If the clump is dense enough to survive the first implosion phase, it comes into a quasi-static equilibrium with the induced pressure \citep{bertoldi90}, in which its radius of curvature is $r_c$ (on the symmetry axis, or the point most directly illuminated by starlight).   Ionization balance in the presence of this accelerating and diverging photoionized flow implies 
\[
{S\over 4\pi R_c^2} = \omega \alpha n_{ic}^2 r_c
\] 
where $n_{ic}$ is the ionized gas density just outside the clump IF at this point, $R_c$ is the distance from the clump to the central source, and $\omega=0.1-0.2$  \citep{bertoldi90} 
The parameter which describes the characteristic ionizing radiation-to-gas pressure ratio ($\Xi_i$) for this outflow is therefore 
\begin{equation} \label{eq:Gamma_clump} 
\Gamma_{c} = {\bar e_i\over 2.2 k T c} \left(\omega \alpha R_c S\over 4\pi r^2\right)^{1/2}. 
\end{equation} 
Compared to conditions at a spherical IF of radius $R_{\rm IF}$, 
\begin{equation}\label{eq:Gamma_clump-to-region} 
 {\Gamma_c^2 \over \Gamma_{\rm IF}^2 } = 3\omega {r_c R_{\rm IF}\over R_c^2}. 
\end{equation} 
A clump generally has a smaller value of $\Gamma$, and therefore $\calU$, than its parent region, because its radius of curvature $r_c$ is significantly less than the ionization front radius $R_{\rm IF}$ of the region.   

A clump's photoevaporation flow ends at a deceleration shock caused by the pressure from ionized gas in the parent region, stellar winds, or radiation.   As $\Gamma_c$ increases toward unity, radiation pressure requires that the shock approach the sonic point of the photoevaporation flow, and when $\Gamma_c >1$, radiation impedes the outflow of ionized gas.  For $\Gamma_c\gg 1$, the ionized gas is confined by radiation pressure to a thin layer on the surface of the clump, leading to a saturation of $\Xi_{i,{\rm em}}$ and $\calU_{\rm obs}$, exactly as described in \S~\ref{SS:RadConfinement} and Appendix \ref{Appendix:PlanarRegions}.   Just as before, dust opacity will be significant so long as $\gamma>1$.  \citet{dopita02} sketch the flow for $\Gamma_c\lesssim 1$. 

However there is one significant difference between the effects of clumpiness and stellar winds. The ionized layer of a clump cannot be hydrostatic:  it must either evaporate away, or be dragged along the clump surface by the tangential component of the radiation momentum flux, and also possibly by the stress from a turbulent boundary layer between photoionized clump gas and shocked stellar winds.  Such wind and photon-driven flows are a plausible source of supersonic line widths in giant \HII regions, for photoevaporative flows only produce motions of up to about twice the ionized sound speed \citep{bertoldi90}. It is not at all clear that a dynamical equilibrium exists which would prevent a clump with $\Gamma_c>1$ from being dynamically disrupted, although a magnetic buffer zone may allow this \citep[e.g.,][]{dursi07}.

\subsection{Magnetic fields} \label{SS:Magnetic} 
Magnetic fields supply an additional form of pressure which is available to balance the force of radiation, and therefore have the potential to raise $\calU_{\rm obs}$ {\em above} $\calU_{\rm obs, max}$.  However this is somewhat difficult to arrange. 

Consider a planar shock of speed $v_s$ running into a medium of hydrogen density $n_0$, magnetic field perpendicular to the shock normal $B_{\perp0}$ and Alfv\'en speed $v_{A0} = B_{\perp0}/(4\pi \mu_H n_0)^{1/2}$.  The post-shocked fluid is in contact with (and driven forward by the pressure of) an ionized layer.   Ignoring any change in pressure across the post-shocked layer as well as streaming of the ionized gas (both of which can be order-unity effects), the density of ionized gas at the contact discontinuity is $(v_s/v_{A0})^2 n_0$ if it is pressure-supported, and $\sqrt{2} (v_s/v_{A0}) n_0$ if it is magnetically supported.  Magnetic pressure support therefore requires 
\begin{equation}\label{eq:Planar_MagneticDominated} 
v_s v_{A0} > \sqrt{2} c_i^2 
\end{equation} 
where $c_i\simeq 10$\,km\,s$^{-1}$ is the ionized sound speed.  Magnetic support is therefore only relevant for shock speeds in excess of about $35 (n_0/100\, {\rm cm}^{-3})^{1/2} T_{i,4} (30 \mu{\rm G}/B_{\perp0})$\,km\,s$^{-1}$.   

Magnetic fields should be relatively more important within clumpy regions, because of the tendency for streaming and turbulent flows to amplify them toward equipartition with the kinetic energy.  

\subsection{Inertially-Confined and Impulsive Regions}\label{InertialConfinement} 

Our discussion so far has focused on regions which are in force balance because ionized gas resides within them for long enough to be crossed by sound waves.   However, some regions will be too young (or have crossing times too long) to develop into this state, such as the regions created in cosmological reionization.  In others, ionized gas flows through the region at speeds well in excess of the ionized sound speed; these include ionized winds and jets,
 ionized ejecta, ionized zones of supersonic turbulence, and \HII regions around runaway stars.   Indeed, champagne flows \citep{whitworth79} are density-bounded to one side and are only approximately in force balance on the side toward the ionization front.  None of these impulsive or inertially-confined flows is limited to $\Xi\lesssim 1$ and $\calU_{\rm obs} < \calU_{\rm obs, max}$ for the reasons we have outlined: therefore, higher values of $\calU_{\rm obs}$ could be taken as spectral evidence of this state. 

 Loosely speaking, an observation of $\calU_{\rm obs}$ above $\calU_{\rm obs, max}$ implies that either a non-thermal form of pressure, such as ram pressure or magnetic fields, balances radiation pressure, that another force (such as gravity) is at play, or that there has not been time to establish equilibrium. 

\subsection{Additional Effects:  
Recombination Line Pressure and Photoelectric Heating}\label{AdditionalEffects}
It is worth noting a couple additional effects involving the interaction among light gas, and dust: 

{\em Lyman-$\alpha$ trapping and dust absorption}. 
In the absence of dust, photons from the Ly-$\alpha$ line scatter many times before escaping.   Once the scattered line radiation is sufficiently strong its pressure will augment gas pressure, leading to an {\em increase} of $\calU_{\rm obs, max}$ \citep [See][for an analysis of line pressure in the AGN context]{elitzur86}. The effect is analogous to the role of magnetic fields, except that the line pressure is proportional to the ionizing radiation pressure.   At solar metallicity, a typical grain population limits the Ly-$\alpha$ line pressure within the ionized zone to only $\sim 6\%$ of the ionized gas pressure \citep{henney98}, but line pressure is significant at low metallicity or in regions cleared of dust.    Even in the absence of dust, the dynamical effect of line pressure limits $\log_{10} \calU\lesssim-1$ within quasi-static regions \citep{1972ApJ...178..105W,ferland84}.  

{\em Dust photoelectric effect}.  Photoelectrons ejected from dust grains represent an additional sources of heat and ionization, while collisions and recombinations  with dust grains are (less significant) sinks.   These processes, considered in detail by \citet{2004MNRAS.350.1330V}, \citet{groves04a}, and others, are fully accounted for within the Cloudy code.  They depend both on the grain abundance (captured here by $\gamma$) and the grain size distribution.  

Of these, trapped line radiation tends to make $\calU_{\rm obs, max}$ to decrease as $\gamma$ increases, whereas photoelectric heating has the opposite effect by raising the ionized gas temperature in dustier gas. 					
																									 
\begin{deluxetable}{cccc}
\tablewidth{0pt}
\tablecaption{Mid-Infrared Forbidden Lines\tablenotemark{a}}
\tablehead{
\colhead{Species} & 
\colhead{$\lambda$ ($\mu$m)} &
\colhead{IP (eV)} &
\colhead{$n_{\rm crit}$ (cm$^{-3}$)}}
\startdata
$\rm  [Ar\,{\footnotesize II}]$  & 6.99 & 15.76 & $2.0\times10^5$ \\
$\rm  [Ar\,{\footnotesize III}]$ & 8.99 & 27.63 & 3.0$\times10^5$ \\
$\rm  [Ne\,{\footnotesize II}]$  & 12.81 & 21.56 & 6.5$\times$10$^5$ \\
$\rm  [Ne\,{\footnotesize III}] $& 15.55 & 40.96 & 1.3$\times$10$^5$ \\
$\rm  [S\,{\footnotesize III}]] $ & 18.71 & 23.34 & 2.0$\times$10$^4$ \\
$\rm [\SIV]$   & 10.51 & 34.79 & 6.0$\times$10$^4$ 
\enddata
\tablenotetext{a}{Values taken from \citet{dopitabook}.}
\label{MIRlines}
\end{deluxetable}

\section{Dust-bounded regions and saturation of $\calU$ from selective dust attenuation} \label{SS:Dust_abs}

Emission lines usually arise from regions large enough to encompass many local environments, so their ratios reflect an average of the  local conditions.   Dust absorption has the effect of dimming high-$\calU$ regions, suppressing their contribution to the average and causing a saturation of the apparent ionization level.   This bias arises from the dust absorption of ionizing radiation within \HII regions, and from the absorption of the forbidden lines themselves, predominantly outside the regions themselves.   

\subsection{Internal dust extinction} \label{SSS:Int_abs}

If a region is sufficiently dusty, then the dust optical depth will exceed unity before radiation pressure becomes important.  The critical value of the dust discriminant  $\gamma$ is apparent if we combine its definition (eq.\ \ref{eq:gamma_in_Dr11}) with the characteristic ionizing radiation-to-gas pressure ratio in a classical Str\"omgren sphere (Eq.\ \ref{eq:Xiem_vs_Gamma_in_Classical}): 
\begin{equation}\label{eq:XiEm_taud_gamma}  
{\tau_{d,{\rm St}} \over \Xi_{i,{\rm em} }} = {4\gamma \over 3}.  
\end{equation}
Although this refers to the classical case of a filled Str\"omgren sphere, for which both $\tau_{d,{\rm St}}\ll1$ and $\Xi_{i,\rm em}\ll1$, it demonstrates that, when $\gamma \gtrsim 3/4$, dust extinction of ionizing photons is significant even for some regions which are not radiation dominated.  Conversely, when $\gamma \lesssim 3/4$, radiation pressure confinement can occur when dust extinction is not strong.     The former scenario (dust-limited regions) is consistent with our expectations of Milky Way conditions, so long as dust is not cleared from the gas; Figure \ref{fig:ParameterSpace} maps the physical regimes in this case.  The latter applies to low-metallicity environments, or more broadly if dust clearing is effective.    (We will not distinguish the critical $\gamma$ from unity in our following discussion.) 

Because recombination light is always redder than the photons which ionize an \HII region, and because dust opacity increases with frequency, the line photons of interest tend to escape the ionized gas relatively unabsorbed.  Nevertheless, because dust consumes ionizing photons, the rate of recombinations is reduced in dust-dominated regions.  The ionization efficiency $f_{\rm ion}$ drops from unity (for a classical \HII region) to an asymptotic value of order $(1+\beta)\tau_d/(\gamma-1)$ (in the planar limit, when $\gamma>1$: Appendix \ref{Appendix:PlanarRegions}) as radiation pressure and dust extinction become significant; see Dr11's Eq.\ (21).  For the parameters of interest, the minimum value of $f_{\rm ion}$ is about 0.3.  

Because of this, recombination light from the regions of highest $\calU_{\rm obs}$ is suppressed by the factor $f_{\rm ion}$ -- so that, on a galactic scale, $\calU_{\rm obs}$ is reduced relative to $\calU_{\rm obs, max}$.  Specifically, so long as the extinction of line photons (\S \ref{SSS:Ext_abs}) is negligible, the inferred value of $\calU_{\rm obs}$ on galactic scales will be approximately the emission-weighted average of all the regions encompassed by the observation: we call this $\calUgal$.   If we label a population of $\cal N$ regions by their ionizing luminosities and apparent ionization parameters, then 
\begin{equation}\label{eq:Uobs_of_population} 
\calUgal= {\int \calU_{\rm obs} f_{\rm ion} S d{\cal N} \over  \int f_{\rm ion} S d{\cal N} }= {\int S^{\xi} \left<f_{\rm ion} \calU_{\rm obs}\right>_S d\ln S  \over \int S^{\xi} \left<f_{\rm ion} \right>_S d\ln S  },
\end{equation}
where in the second equality we assume $d{\cal N}/d\ln S \propto S^{\xi-1}$ and define $\left<x\right>_S$ to be the average of $x$ among regions whose ionization rates lie in a narrow range around $S$.    Galactic \HII region luminosity functions obey $\xi\simeq 0$ at high $S$, but are significantly shallower at lower luminosities \citep[$S \lesssim 10^{50}$\,s$^{-1}$, where the IMF sampling is incomplete;][]{1989ApJ...337..761K}.  Each decade of $S$ above this break therefore contributes about equally to $\calUgal$, while those below it are increasingly insignificant. 

In general, a galaxy with more massive star clusters (a population extending to higher $S$) and a denser circum-cluster medium (so that regions are more compact) will have regions characterized by higher values of $\calU_{\rm obs}$ and therefore attain a higher $\calUgal$.  However, the decline in $f_{\rm ion}$ accompanying the increase in $\tau_{d,{\rm St}}$ suppresses the contribution of those regions with highest $\calU_{\rm obs}$. 
So long as the \HII region population includes a range of intensities and a significant range of $\tau_d$ and $f_{\rm ion}$,  
the net effect will be that $\calUgal$ is an underrepresentation of the starlight-weighted average of $\calU_{\rm obs}$ (the result of taking $f_{\rm ion}\rightarrow 1$ in eq. \ref{eq:Uobs_of_population}).  In particular, $\calUgal$ will be closer to the value it has when $\tau_{d, {\rm St}} = 1$,
\begin{equation} \label{eq:calU_for_taud=1}
 {3\over 4\gamma} {2.2 k_B T_i\over \bar e_i} = 10^{-2.25} {10\over \gamma}
{T_4\over 0.8}  {20\,{\rm eV}\over \bar e_i}. 
\end{equation} 

It is important to remember that this selection effect does not apply to individual \HII regions, or parts thereof, which saturate at a higher value ($\calU_{\rm obs, max}$, \S \ref{SS:RadConfinement}).   It operates only when starlight photons are distributed rather evenly across environments with a range of $\calU_{\rm obs}$ and $f_{\rm ion}$, so that variations of $f_{\rm ion}$ can affect the apparent value of $\calU_{\rm obs}$.   For this reason, it does not prevent $\calUgal$ from limiting $\calU_{\rm obs, max}$ if {\em all} the \HII regions in a galaxy are radiation-confined.  

\begin{figure}
\epsscale{1}
\plotone{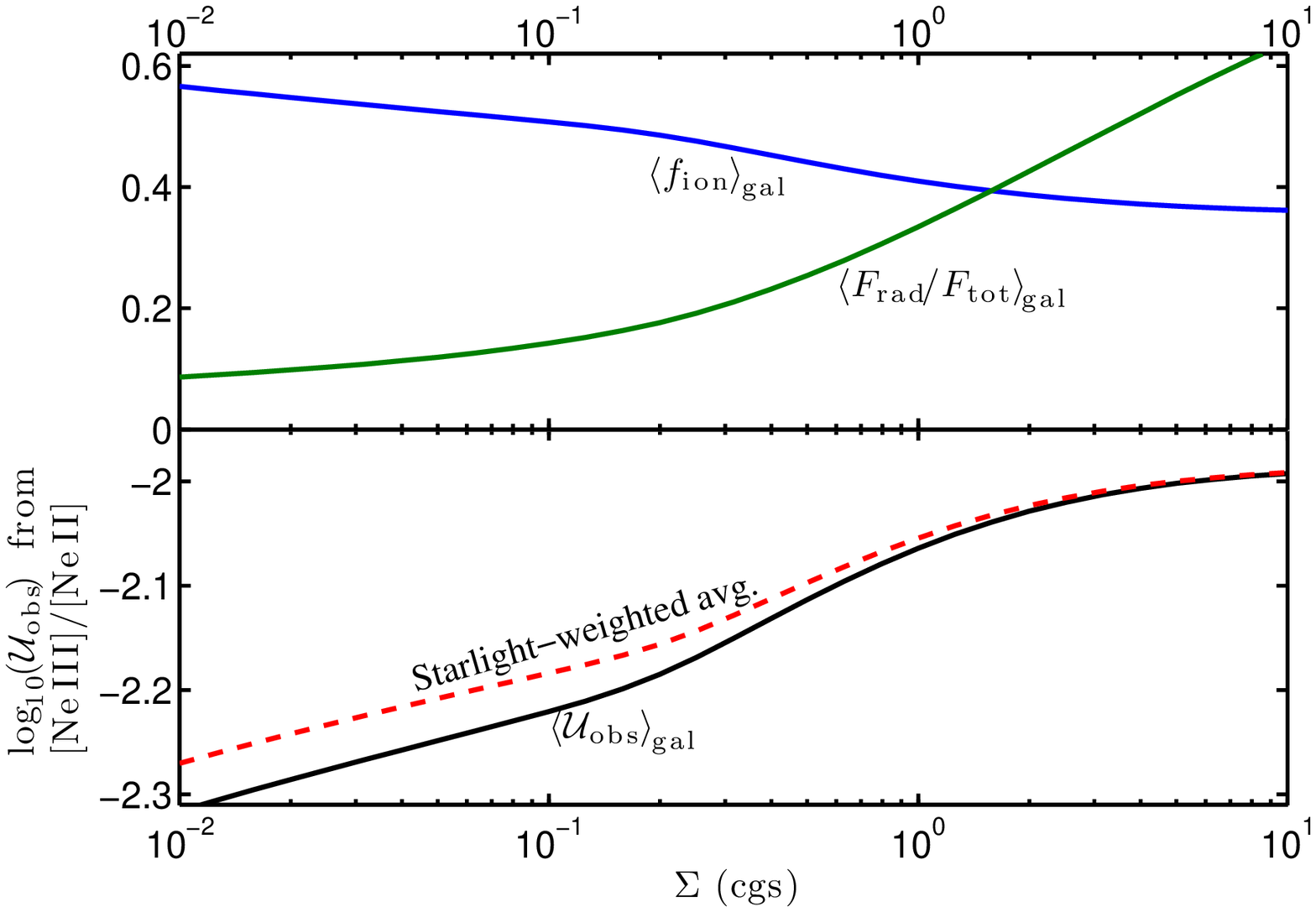}
\caption{Influence of internal dust absorption on the fraction of ionizing photons caught by H atoms rather than dust ($\left<f_{\rm ion}\right>_{\rm gal}$, top panel), galaxy-averaged ratio of radiation to total (gas+radiation) force ($\left<F_{\rm rad}/F_{\rm tot}\right>_{\rm gal}$, top panel) and apparent ionization parameter ($\calUgal$, bottom panel), in a toy model for a galactic population of \HII regions (\S \ref{SSS:Int_abs}).  For comparison we also plot the starlight-averaged value of $\calU_{\rm obs}$ (eq.\ \ref{eq:Uobs_of_population} with $f_{\rm ion}\rightarrow 1$).  In this model, the luminosity function and lifetimes of the driving associations are derived from Monte-Carlo modeling by sampling the stellar IMF \citep{matzner02}.  Regions expand due to a combination of radiation and photoionized gas pressures \citep{km09} in a medium of definite column density $\Sigma$, and are taken to absorb and emit light as in the numerical models of \S \ref{S:model}.   }
\label{fig:UobsgalModel}
\end{figure}

We demonstrate these points in Figure \ref{fig:UobsgalModel}, for which we generate a toy model of the \HII region distribution within a galaxy.    We start by examining a population of $10^6$ associations generated for a prior work (\citealt{matzner02}, based on \citealt{mw97}, in which $\xi=0$ at high $S$), extracting from it the luminosity function $d{\cal N}/d\ln S$ and the typical ionizing lifetime $\left<t_{\rm ion}\right>_S$ as functions of $S$.   We then realize a scenario in which all \HII regions expand into density profiles with $\rho \propto r^{-1}$, taking the mean column density $\Sigma=2r\rho$ to be constant and the same for all regions.  (For simplicity, we neither vary $\Sigma$ with $S$, nor alter the upper limit of $S$ with $\Sigma$ -- although both effects are likely to occur in real galaxies.)   In each luminosity bin we evaluate the time evolution $\Psi(t)$ using KM09's equation (13), which accounts for the combined force due to radiation, partially trapped winds (if any), and photo-ionized gas.   We limit regions' growth according to their finite initial size and stalling in the finite hydrostatic pressure of the environment (KM09's \S~2.3).  We assume for this model that stellar winds are weak: $\log_{10}\Omega = -1.5$.   With this assumption, we can construct $f_{\rm ion}(S,t)$ and $\calU_{\rm obs}(S,t)$ (from the \NeIII/\NeII\ line ratio) using our own numerical models (\S \ref{S:model}).  Taking the time average for $0<t<\left<t_{\rm ion}\right>_S$ to get $\left<f_{\rm ion}\right>_S$ and  $\left<f_{\rm ion} \calU_{\rm obs}\right>_S$, we arrive at $\calUgal$.   For comparison we also plot the starlight-weighted averages of $\calU_{\rm obs}$,  $f_{\rm ion}$, and the radiation-to-gas force ratio $F_{\rm rad}/F_{\rm tot} = (1+\Psi^{1/2})^{-1}$ (see KM09 eq. 10).   Although the dust selection effect is not strong in our toy model, amounting to a suppression  of $\calUgal$ by only $\sim 7-12\%$ over a relevant range of $\Sigma$, we suspect it might be more significant if some of our toy model assumptions were relaxed.  

Our toy model does not account for heterogeneity of the ambient medium, which permits the existence of a diffuse ionized component; it is likely to under-estimate $\left<f_{\rm ion}\right>_{\rm gal}$ and over-estimate $\calUgal$ for this reason. 

\subsection{External dust extinction} \label{SSS:Ext_abs} 

In the expansion of an \HII region, most of the ambient medium is swept before the ionized gas rather than being ionized itself.   Therefore, although dust within the \HII region does not severely attenuate recombination line photons, the surrounding matter poses a thicker barrier.   This external dust extinction is also an increasing function of $\calU_{\rm obs}$, as the following argument shows. 

Suppose that each star cluster, of mass $M$, forms within an overdensity of radial profile $\rho = \rho_0 r^{-k}$, and moreover that, statistically speaking, $\rho_0\propto M^j$.  The toy model of \S~\ref{SSS:Int_abs} adopted $k=1$ and $j=0$, but $k$ could be as low as 0 or as high as 2, and $j$ is likely to be positive.   For the star clusters massive enough to sample the IMF, $L\propto S\propto M$.   

How then does the dust optical depth of the neutral matter outside an \HII region scale with $\calU$ within it?  We restrict attention to $k>1$, so that optical depth varies significantly with radius. When the region size is $R_{\rm IF}$, the external dust optical depth at any frequency is then $\tau_{d,{\rm ext} }\propto R_{\rm IF} \rho(R_{\rm IF}) = \rho_0 R_{\rm IF}^{1-k}$; at the same time, the radiation pressure parameter $\Psi\propto R_{\rm IF}/S$, so that 
\begin{equation} \label{eq:tau_vs_Gamma_regions} 
\tau_{d,{\rm ext}} \propto S^{j+1-k} \Psi^{1-k}
\end{equation} 
for massive clusters.  When $k>1$, regions of a given $S$ are most enshrouded when they are most radiation-dominated (i.e., when $\Psi$ is lowest).   The region luminosity increases or decreases $\tau_d$ at {\em fixed} $\Psi$, depending on whether $j>k-1$ or not.   However, regions with higher $S$ tend to also be more radiation dominated (i.e., have lower $\Psi$).   When $j>k-1$ it is clear that $\tau_d$ decreases as $\Psi$ increases.  Insofar as this external dust is optically thick at the frequencies of the lines being observed, this will tend to suppress emission from high-$\calU_{\rm obs}$ regions, reducing this quantity relative to its theoretical maximum value.   Note that, as we saw in \S~\ref{SS:RadConfinement}, the radiation dominated state which produces $\calU_{\rm obs, max}$ is in fact impossible if the column density of the ambient medium is too low.  

The external attenuation of line photons is minimized by considering the longest possible transition wavelengths; we focus  on mid-infrared lines in \S~\ref{S:model}.   It should also be mitigated by holes in the swept-up neutral shell, as well as clumpiness of the ambient dust distribution (e.g., \citealt{fischera05}). 

\section{Numerical models of quasi-static \HII regions}\label{S:model}

In this section we present numerical results for spherical, dusty, quasi-static \HII regions.    Our goal is to account explicitly for the possibility of an inner boundary pressure due to shocked wind (as parameterized by $\Omega$)  and to demonstrate the saturation of line ratios and inferred $\calU_{\rm obs}$ due to radiation pressure confinement.   
 We consider only a few ionizing spectra and do not vary the composition of gas or the  grain population; a full exploration of these parameters is beyond the scope of this work. 

We use Cloudy version 08.00, last described by \citet{cloudy98}, to account for many important microphysical effects including the photoelectric effect, collisional cooling, and the pressure due to optically thick recombination lines.   However we consider only quasi-static regions in perfect force balance, and do not account for secular effects, such as grain drift, which might lead to inhomogeneities in the composition. 

\subsection{Spectral synthesis and photoionization models}
Using Starburst99 \citep{sb99}, we generate the ionizing continua from coeval star clusters of different ages, all of which we assume are massive enough to fully sample the stellar initial mass function, which we take to have exponents -1.3 and -2.3 between stellar mass boundaries 0.1, 0.5, and 120 M$_\sun$.  We employ the Geneva high mass-loss evolutionary tracks with solar metallicity. These are optimized 
for modeling atmospheres of high mass stars and are recommended by \citet{mm94}.  We adopt Pauldrach/Hillier atmospheres, as these include non-LTE and line-blanketing effects \citep{smith02} for O stars \citep{pauldrach01} and Wolf-Rayet stars \citep{hm98}. The combination of the Geneva high mass-loss tracks and Pauldrach/Hillier atmosphere is recommended when Wolf-Rayet stars are important \citep{vazquez05}.   Starburst99 output spectra are recorded from 0 to 11 Myr with 0.5 Myr steps.

Starburst99 output continuum spectra are fed into Cloudy as the ionizing continuum of each simulated \HII region.
Each \HII region is spherical and in perfect force balance;
we allow radiation pressure to exceed gas pressure, in contrast to Cloudy's default setting. 
We adopt Cloudy's default ISM abundances and dust grain size distributions. 
Each calculation stops where temperature drops to 100 K, and so encapsulates the IF. 
Each set of the simulations outputs the integrated luminosity of selected MIR emission lines form  Table~\ref{MIRlines}, 
including [\ArIII]$_{\lambda 9.0\micron}$, [\ArII]$_{\lambda 7.0\micron}$, [\NeIII]$_{\lambda 15.5\micron}$, 
[\NeII]$_{\lambda12.8\micron}$, [\SIII]$_{\lambda 18.7\micron}$, and [\SIV]$_{\lambda 10.5\micron}$.

\subsection{Mapping line ratios to  $\calU_{\rm obs}$} \label{SS:UobsMethod} 

\begin{figure}
\epsscale{1}
\plotone{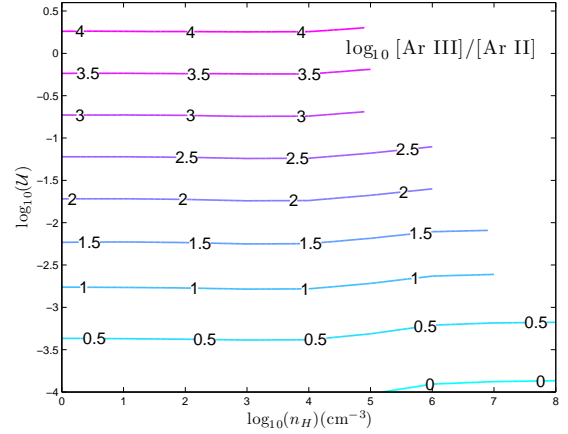}
\caption{Ratio of \ArIII$_{8.99\,\mu{\rm m}}$ to \ArII$_{6.99\,\mu{\rm m}}$ luminosities as a function of $\calU$ and $n_H$ in homogeneous mixtures of solar metallicity gas with the ionizing spectrum of a fully sampled coeval star cluster at 2 Myr age.  The change in line ratio across the critical densities of $2-3\times 10^5$\,cm$^{-3}$ is clearly visible if not dramatic.   }
\label{fig:ar3ar2_1zone}
\end{figure}

\begin{figure}
\epsscale{1}
\plotone{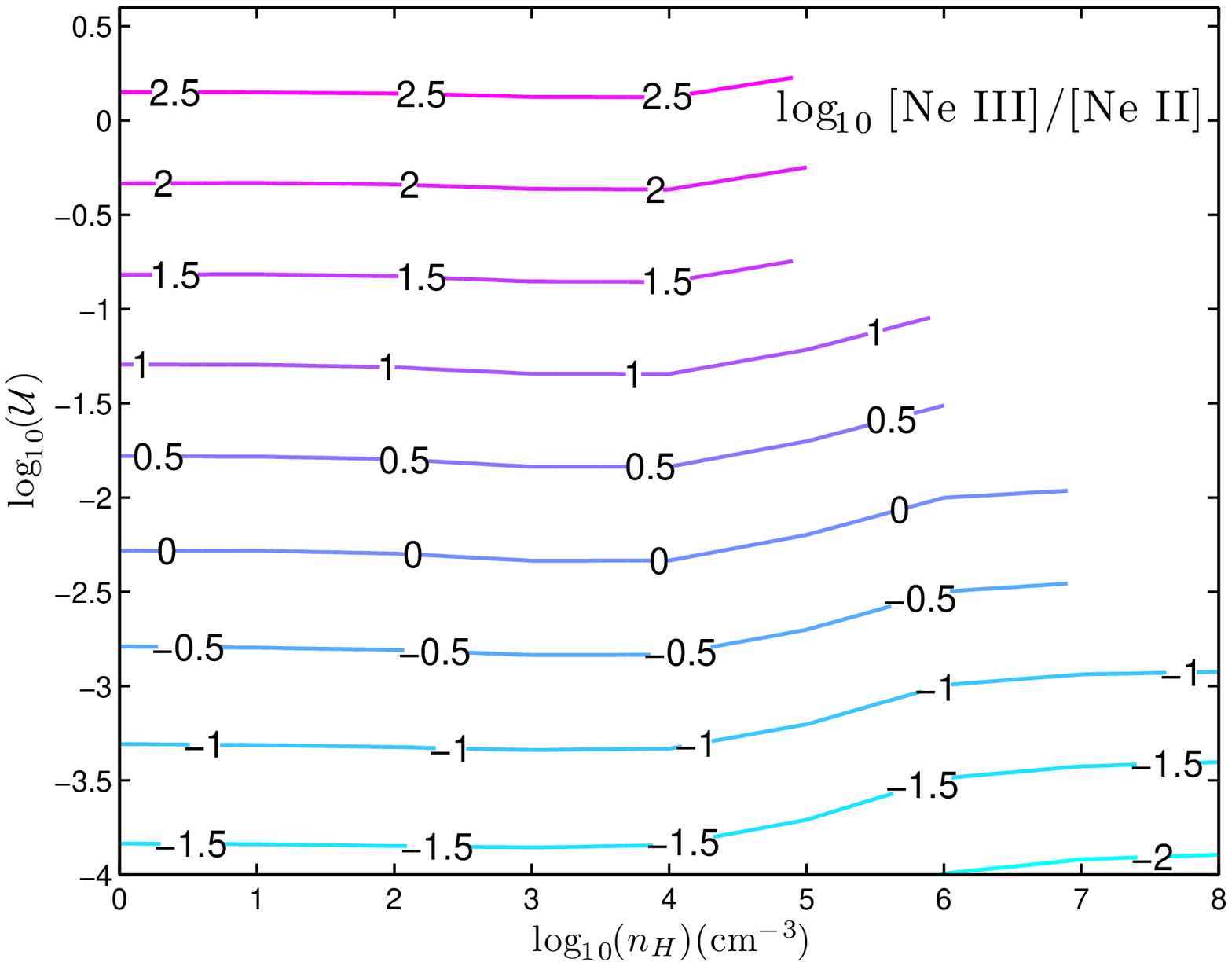}
\caption{As in figure \ref{fig:ar3ar2_1zone}, but for \NeIII$_{15.55\,\mu{\rm m}}$/\NeII$_{12.81\,\mu{\rm m}}$.   }
\label{fig:ne3ne2_1zone}
\end{figure}

\begin{figure}
\epsscale{1}
\plotone{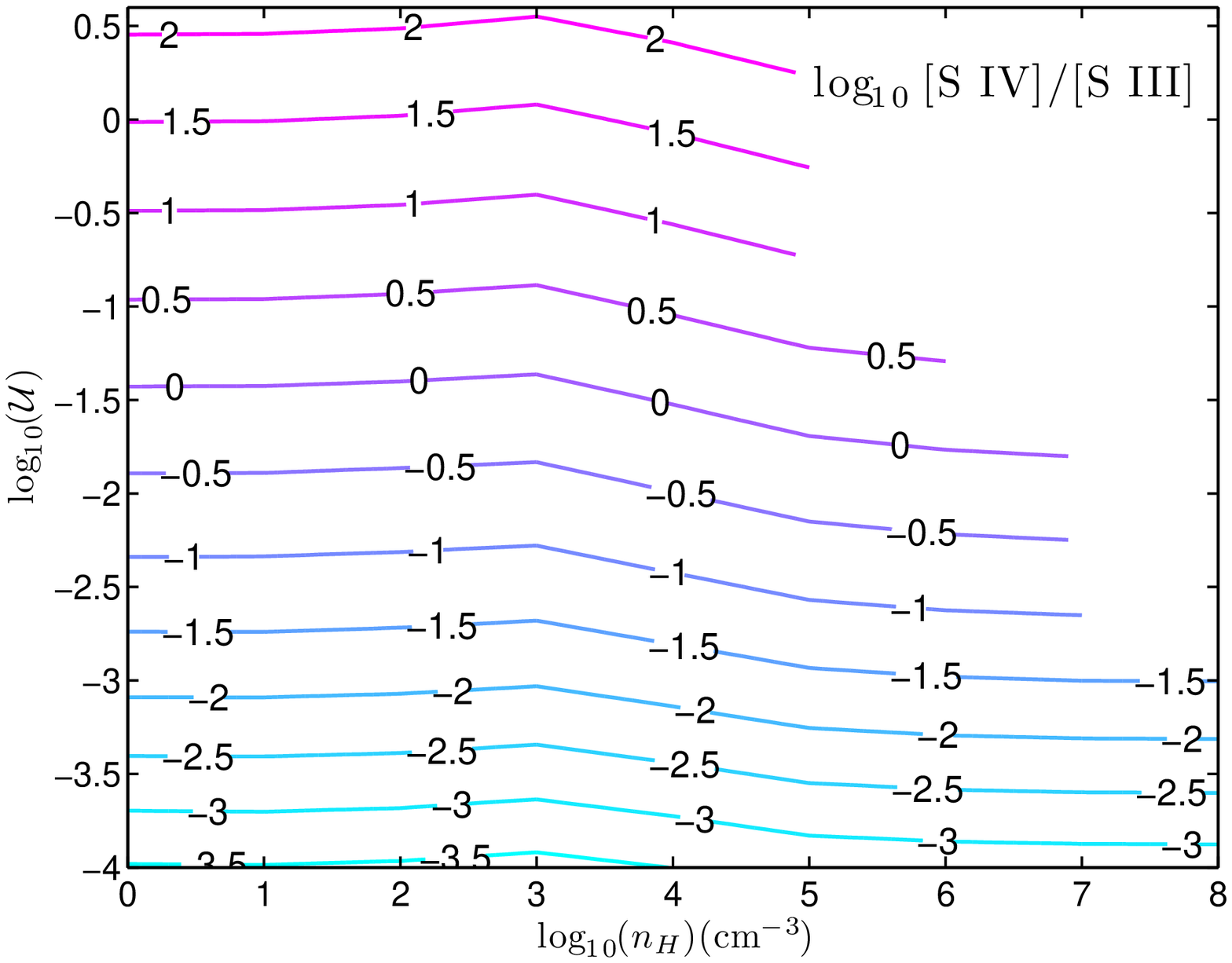}
\caption{As in figure \ref{fig:ar3ar2_1zone}, but for \SIV$_{10.51,\mu{\rm m}}$/\SIII$_{18.71\,\mu{\rm m}}$.   }
\label{fig:s4s3_1zone}
\end{figure}

In both observational and theoretical studies it is necessary to translate some observed line ratios into a set of physical parameters, including $\calU$; as we have already discussed,  we use $\calU_{\rm obs}$ to designate the apparent $\calU$ which characterizes a single region. A standard definition for $\calU_{\rm obs}$ is the value of $\calU$ at the inner boundary of an (often uniform) slab which reproduces the observed lines; however this has pitfalls both for low and high $\calU$ (\S \ref{conclusion}), and ignores the possible role of wind pressurization.  We choose instead to associate $\calU_{\rm obs}$ with the value of $\calU$ within a {\em homogeneous} mixture of gas and radiation.   To do so, we require a library of the line luminosities emitted by such homogeneous mixtures, in which physical parameters such as $\calU$, $n$, and input spectrum are varied.  We construct this library using the very innermost zone of a sequence of Cloudy models.   Selected results are presented in figures \ref{fig:ar3ar2_1zone} to \ref{fig:s4s3_1zone}.  Figure \ref{fig:onezoneU} demonstrates how well the  [\NeIII]/[\NeII] ratio reflects $\calU$. 
Other line ratios,  [\ArIII]$_{\lambda 8.99  \micron}$/[\ArII]$_{\lambda 6.99 \micron}$ 
and  [\SIV]$_{\lambda 10.5 \micron}$/[\SIII]$_{\lambda 18.7 \micron}$, are also most sensitive to the intensity of radiation field, i.e., to $\calU$. 

\subsection{Model parameters} \label{SS:modelparameters} 
We work with a fixed dust population and several input spectra corresponding to fully sampled, coeval star clusters of different ages.  Therefore, while the ratio $\beta$ of non-ionizing to ionizing radiation force varies greatly from young (2 Myr) to relatively old (6 Myr) cluster spectra, the dust discriminant $\gamma$ is almost constant.   We list the relevant quantities in Table \ref{Table:SpectraParameters}.   Whenever we are not exploring age dependence, we employ spectra from 2 Myr clusters, as this is halfway through their typical ionizing lifetimes. 

\begin{deluxetable}{cccccc}
\tablewidth{0pt}
\tablecaption{Coeval Massive Cluster Ionizing Spectra}
\tablehead{
\colhead{Age (Myr) } & 
\colhead{$\bar e_i$ (eV) } &
\colhead{$L_n/L_i$ } & \colhead{$\beta$\tablenotemark{a}} & 
\colhead{$\log_{10} \sigma_d$} &
\colhead{$\gamma$}
}
\startdata
0 & 20.6 & 0.61 & 0.61 &  -20.9& 11.2  \\
2 & 19.1 & 1.25 & 1.12 & -20.8&  13.8 \\
4 & 19.2 & 2.81 & 2.51 & -20.8& 13.2  \\
6 & 16.7 & 16.7 & 11.3 & -20.7& 19.3 
\enddata
\label{Table:SpectraParameters}
\tablenotetext{a}{For $\beta$ we list the ratio of non-ionizing to ionizing radiation {\em force} transferred to the dust grains by an unattenuated input spectrum, differentiating this from $L_n/L_i$, the definition used by Dr11.} 
\end{deluxetable}

For each spectrum our goal is to map the line luminosities, line ratios, and resulting $\calU_{\rm obs}$ as function of $\Psi$ and $\Gamma$, i.e., to scan the parameter space displayed in Figure \ref{fig:ParameterSpace}.   
Practically, we accomplish this by varying the central luminosity and innermost density of the region, at a fixed inner radius.   
For this study we restrict ourselves to regions with ionized densities well below the critical densities for the transitions being considered (Table \ref{MIRlines}) so that density does not enter as a third parameter.   Our plots in this section should therefore not be used for regions with pressures exceeding $10^8\,k_B$\,cm$^{-3}$ for  [\SIII] and  [\SIV], or $10^9\,k_B$\,cm$^{-3}$ for the others.   Moreover we keep the metallicity of gas and stars fixed, and vary only the age of the stellar population (rather than its IMF or other properties).  

In addition to the line luminosities, we compute $f_{\rm ion}$ from the ratio of total H$\beta$ emission to the amount of H$\beta$ emission expected in the absence of dust.  We also compute $S n_{\rm em}$, a useful observational diagnostic of the strength of ionization, where 
\begin{equation}\label{def:n_em}
n_{\rm em} \equiv {\int n^3 dV \over \int n^2 dV},
\end{equation} 
the recombination-averaged density -- a proxy for the observationally inferred density $n_{\rm obs}$. 
 For regions without internal wind bubbles, $S n_{\rm em}$ determines $\Gamma$ and $\tau_d$ (\S \ref{S:Parameters}, Dr11); but pressurization breaks this relation.

\subsection{Results}
\begin{figure}
\epsscale{1}
\plotone{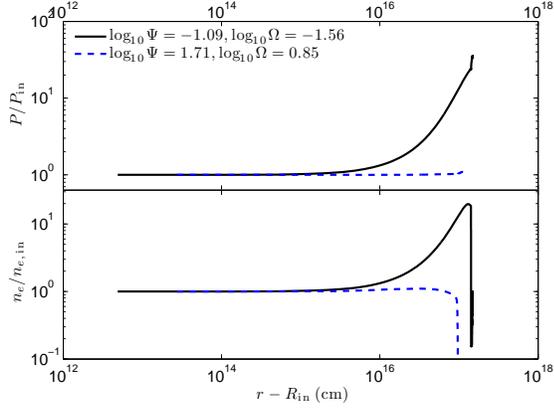}
\caption{Normalized pressure (top) and electron density (bottom) profiles of \HII regions, showing the structure of radiation pressure dominated (solid) and wind bubble (dashed) regimes, which were illuminated by the ionizing spectrum of a 2 Myr cluster of $10^{42}$\,erg\,s$^{-1}$ and $10^{39}$\,erg\,s$^{-1}$  luminosity, respectively.  A drop in temperature beyond the ionization front leads to an increase in density which outpaces the declining ionization fraction, causing an uptick in $n_e$  in the outermost zones.  }\label{fig:ne_P_twomodels}
\end{figure}

\begin{figure}
\epsscale{1}
\plotone{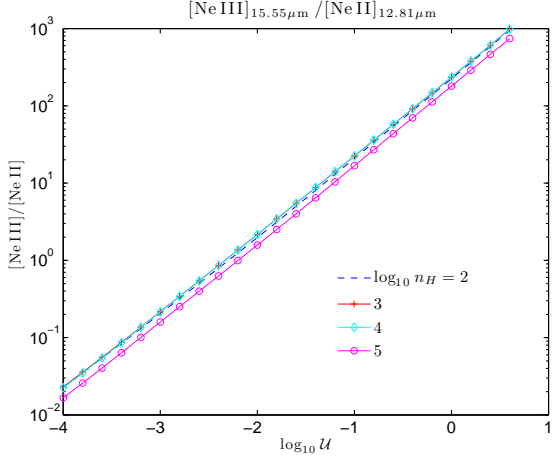}
\caption{One-zone [\NeIII]/[\NeII] line ratio at various values of $\log_{10}\calU$, for 
hydrogen density $\log_{10}n_H = 2, 3, 4, 5$ (blue dashed line, red crossed line, 
cyan line with diamonds, magenta line with circles, respectively). Metallicity and ionizing spectrum are held fixed: $Z_\odot$ and the spectrum of a 2 Myr cluster.  The [\NeIII]/[\NeII] ratio is sensitive to $\calU$ and insensitive to $n_H$, falling only 37\% for densities which approach the critical density of the [\NeII] line. 
}\label{fig:onezoneU}
\end{figure}

We begin by comparing the physical structures of a radiation-confined shell ($\log_{10}\Psi = -1.09$, $\log_{10} \Omega = -1.56$) with those of a wind-confined shell ($\log_{10}\Psi = 1.71$, $\log_{10} \Omega = 0.85$) in Figure \ref{fig:ne_P_twomodels}.   As expected, radiation confinement leads to a dramatic gradient of electron density and pressure, whereas wind pressure causes the region to be nearly uniform.

Figures~\ref{fig:ne3ne2_2myr} to \ref{fig:ne3ne2_6myr} show 
$\calU_{\rm obs}$ derived from [\NeIII]/[\NeII] ratios for cluster ages 2, 4, and 6 Myr. 
Immediately apparent is the saturation of $\calU_{\rm obs}$ in the radiation-confined state,  $\log_{10}\Psi  < 0$ and 
$\log_{10} \Omega < 0$, as is its suppression in wind-confined shells ($\log_{10}\Omega >0$). 

At age 4 Myr, the most massive stars 
have left the main sequence, so the ionizing flux drops and $\calU_{\rm obs}$ values decrease accordingly.  The pattern of saturation in the radiation-dominated quadrant is, however, unchanged. 

By 6 Myr, only stars with mass $\leq$ 30 M$_\sun$ are still on the main sequence, which are responsible for producing
only 7.5\% of the ionizing flux at zero age. Therefore $\calU_{\rm obs}$ values decreased gradually comparing to that at younger ages.
Note that $\calU_{\rm obs}$ still saturates in the same manner as the 2 Myr and 4 Myr cases. 

A second set of contours in the $\Psi$--$\Omega$ plots, labeled with dashed black lines, indicate $\log_{10}(Sn_{\rm em})$. 
This quantity increases
as $\Psi$ decreases, which implies that dust opacity increases as radiation pressure also becomes more important. This is consistent 
with the analytical prediction in \S~\ref{S:Parameters}. $Sn_{\rm em}$ values are 
weakly correlated with $\Omega$ in the $\log_{10}\Omega < 1$ regimes where stellar winds are negligible. Then the contours 
become a linear function of both $\Psi$ and $\Omega$ when $\log_{10}\Omega > 1$. 

As we have discussed in \S~\ref{SS:Dust_abs}, dust grains absorb a fraction $1-f_{\rm ion}$ of the ionizing photons in \HII regions. 
We estimate  $f_{\rm ion}$  using the ratio of total H$\beta$ emission to the dust-free value H$\beta_0$, which we calculate by assuming that all ionizing photons are used for photoionization, and each ionizing photon is responsible for one recombination. 
The emitted H$\beta$ luminosity -- which we tally zone-by-zone to avoid extinction of the line photons --  
 reflects the suppression of ionizations due to dust absorption of starlight. 
The variation of H$\beta$/H$\beta_0$ with $\Psi$  and $\Omega$
is shown in Figure~\ref{fig:Fion} with solid blue contours, for the ionizing spectrum of a 2 Myr-old cluster. Again, dashed black contours denote $\log_{10} (Sn_{\rm em})$, 
providing a calibration with respect to previous $\calU_{\rm obs}$ contour plots at the same age. 
In the quadrant corresponding to radiation-confined shells, the H$\beta$/H$\beta_0$ line ratio is 0.35. This is consistent with our prediction, in Appendix \ref{Appendix:PlanarRegions}, that $f_{\rm ion}$ falls to $\sim 0.3$ in the limit $\Psi\rightarrow 0$, $\Omega\rightarrow 0$ for the values of $\gamma$ and $\beta$ which characterize a 2 Myr cluster.

\section{Discussion: implications of $\calU_{\rm obs}$ saturation in starburst galaxies}\label{S:implication}

\begin{deluxetable}{ccccc}
\tablewidth{0pt}
\tablecaption{Physical parameters of starburst galaxies}
\tablehead{
\colhead{Object} & 
\colhead{$n_e$ (cm$^{-3}$)} &
\colhead{$\log_{10}S$ (s$^{-1}$)} &
\colhead{$\calU_{\rm obs}$} &
\colhead{References}}
\startdata
M82 central 500 pc & 10 -- 600 & 54.09 & -2.1 -- -2.6 & a \\
NGC 253  & 430$^{\rm{+290}}_{\rm{-225}}$ & 53.00 & -2.2 -- -2.6 & b, c, d \\
NGC 3256 & $\leq$1400 & 52.30 -- 53.78 & -2.3 & e \\
Antennae & \nodata & \nodata & -1.57 & f 
\enddata
\tablenotetext{a}{\citet{fs01}}
\tablenotetext{b}{\citet{carral94}}
\tablenotetext{c}{\citet{engel98}}
\tablenotetext{d}{\citet{thornley00}}
\tablenotetext{e}{\citet{roy05}}
\tablenotetext{f}{\citet{snijders07}}
\label{Table:physical_quantity}
\end{deluxetable}
 
On the basis of the theory developed above, we expect to find that individual \HII regions (or regions within them) display line ratios consistent with ionization parameters ranging up to the maximum value  corresponding to radiation confinement.  Although the same physical process limits the inferred ionization parameter for observations on galactic scales, the value of the upper limit is not the same, because of the different measures used: usually geometric estimates (our $\calU_{\rm geom}$) in resolved regions, as opposed to pure line ratios (our $\calU_{\rm obs}$, or more precisely $\calUgal$) in unresolved ones.   Moreover, averaging over many regions tends to blend together those at $\calU_{\rm obs, max}$ with those below it, and the selective dimming of high-$\calU_{\rm obs}$ regions by internal dust absorption favors those below the maximum. 

\begin{figure}
\epsscale{1}
\plotone{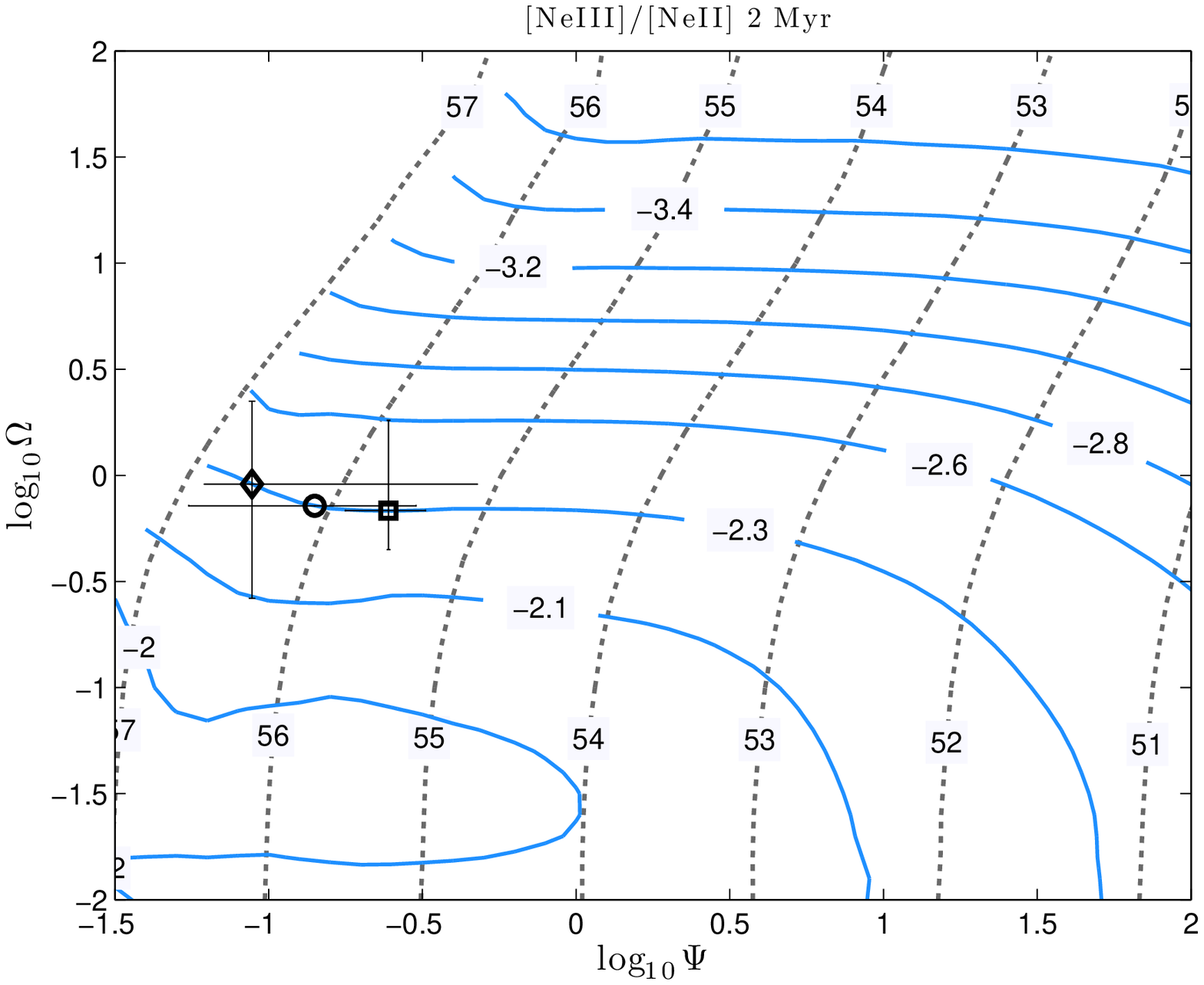}
\caption{Logarithm of ionization parameter (solid blue contours) as a function of $\log_{10}\Psi$ and $\log_{10}\Omega$, for illumination by a 2 Myr-old cluster. The $\log_{10}\calU_{\rm obs}$ values are evaluated from the line ratio  [\NeIII]$_{\lambda 15.55 \micron}$/[\NeII]$_{\lambda 12.81 \micron}$. Dashed black contours denote the logarithm of $Sn_{em}$. \HII region of the central 500 pc of M82, NGC 3256, and NGC 253, are marked by open diamond, circle, and square, respectively.  The maximum value corresponds to $\calU_{\rm obs, max} = -1.97$. }
\label{fig:ne3ne2_2myr}
\end{figure}

\begin{figure}
\epsscale{1}
\plotone{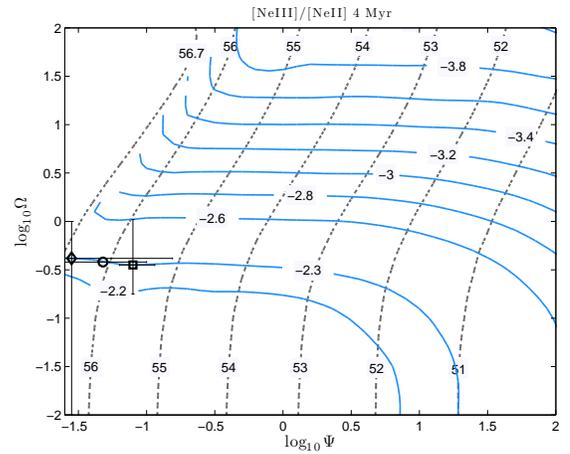}
\caption{Same as in Figure~\ref{fig:ne3ne2_2myr}, but for ionization by a star cluster of age 4 Myr.  The $\log_{10}\calU_{\rm obs}$ values are evaluated from the line ratio  [\NeIII]$_{\lambda 15.55 \micron}$/[\NeII]$_{\lambda 12.81 \micron}$. Symbols that mark \HII regions of M82, NGC 3256, and NGC 253, are the same as in Figure \ref{fig:ne3ne2_2myr}.
}\label{fig:ne3ne2_4myr}
\end{figure}

\begin{figure}
\epsscale{1}
\plotone{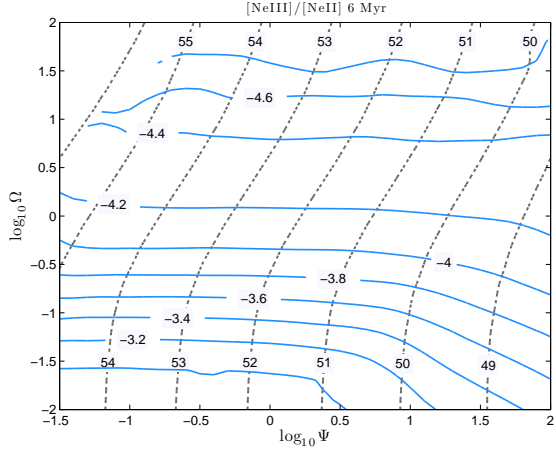}
\caption{As in Figure~\ref{fig:ne3ne2_2myr}, for ionization by a star cluster of 6 Myr age. Values of $\log_{10} \calU_{\rm obs}$ are evaluated from the line ratio  [\NeIII]$_{\lambda 15.55 \micron}$/[\NeII]$_{\lambda 12.81 \micron}$.
}\label{fig:ne3ne2_6myr}
\end{figure}

\begin{figure}
\epsscale{1}
\plotone{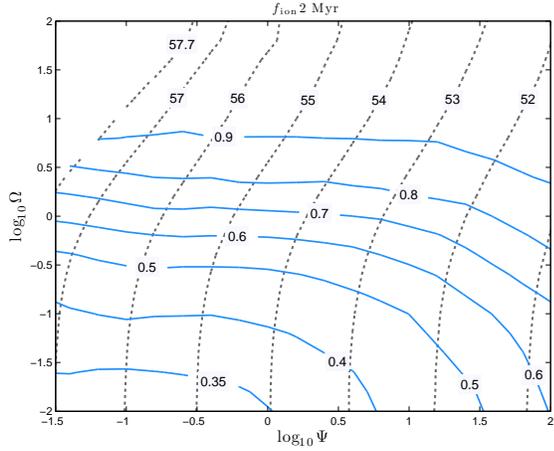}
\caption{Solid blue contours mark $f_{\rm ion}$, calculated as the ratio of the H$\beta$ line to H$\beta_0$, its value in the absence of dust, as a function of $\log_{10}\Psi$ and $\log_{10}\Omega$. Dashed black contours label $\log_{10}(Sn_{\rm em})$ as in previous figures.  The ionizing spectrum corresponds to a star cluster age of 2 Myr.
}\label{fig:Fion}
\end{figure}

We estimate the upper limit of $\calU_{\rm geom}$ to be roughly $10^{-1}$ using the solutions of Appendix \ref{Appendix:PlanarRegions}, by replacing $n_{i,{\rm obs}}$ with the emission-weighted density $n_{\rm em}$.  This is only approximate, as it does not account for the formation of the lines used to derive $n_{i,{\rm obs}}$.   For unresolved regions we apply $\calU_{\rm obs, max}$, in which we do account for line formation: because most of the observations employ the the [\NeIII]/[\NeII] line ratio of figure \ref{fig:ne3ne2_2myr}, we apply the upper limit of $10^{-1.97}$ we found there.   Part of the difference between the two upper limits can be seen in the internal structure of radiation-confined zones, in which $\Xi_{\rm geom}$ is about 0.3 dex higher than the recombination-weighted average $\Xi_{i,{\rm em}}$.   The rest of the difference is presumably due to the details of line formation. 

To test these expectations against observed regions, we plot in Figure  \ref{fig:Uexamples} a sample of observationally-derived ionization parameters from the literature, divided according to the scope of the observations involved and according to the method used.   As expected, the data range up to values which are approximately our estimated maximum value for each method.  For individually-resolved regions, the observed upper limit appears to be somewhat higher than our prediction, but we attribute this to the approximate nature of our $\calU_{\rm geom, max}$ (rather than to magnetic fields, say).   For galactic-scale observations highest $\calU_{\rm obs}$ is slightly lower than our $\calU_{\rm obs, max}$, which could be a product of the averaging process.  

These results in hand, we turn to the importance of wind pressure in \HII regions. 

For a resolved, spherical \HII region with well-constrained observations of $S$, $n_e$, and $\calU_{\rm obs}$, it is possible to estimate the wind parameter $\Omega$ by reference to a plot like our Figure \ref{fig:ne3ne2_2myr}.   
With only $\calU_{\rm obs}$ or $\calU_{\rm geom}$, one obtains an upper limit on $\Omega$. 
By this logic, we can state quite firmly that the wind parameter $\Omega$ is often small in resolved \HII regions: otherwise, our example values of $\calU_{\rm obs}$ in the upper panel of Figure \ref{fig:Uexamples} would be much less than the theoretical maximum. 

The same argument applies, if somewhat more weakly, to entire galaxies:  for them, $\log_{10} \calU_{\rm obs}$ ranges up to about  $-2.3$.  This is far higher than what would be possible if they were adiabatic wind bubbles (Eq.\ \ref{UgeomInWeaver77}) unless they are exceptionally weak.   This can be seen, for instance, in Figure \ref{fig:ne3ne2_2myr}, where we use the values of $\calUgal$ and the product of {\em total} $S$ and inferred $n_{i, {\rm obs}}$ to place several starburst galaxies within the parameter space of individual \HII regions.    Because this is an upper limit on the actual values of  $Sn_{\rm em}$ within each galaxy's \HII regions, and because $\calUgal$ cannot be higher than the typical values of $\calU_{\rm obs}$ in those regions,  we infer that the characteristic values of $\log_{10} \Omega$ cannot exceed about $0.25$. 

While wind-dominated bubbles are likely to exist, we find no evidence for them either in individual regions or on galactic scales.  We conclude from this that the leakage of wind energy past the IF, either by advection or radiation, is significant.

\section{Conclusions}\label{conclusion}
Our primary motivation in this study has been to assess whether the ionization parameter can be used to measure the importance of radiation pressure in the inflation of \HII regions in external galaxies, especially starburst galaxies.  We have been interested in how well the observational estimate $\calU_{\rm obs}$ reflects the radiation-to-gas pressure ratio $\Xi$ in local conditions and its average over galactic scales.   Because $\Xi\propto \calU$ in each parcel of gas, and because $\calU$ controls the ionization state and therefore the emitted spectrum, it is possible, in principle, to derive the role of photon momentum in the kinematic feedback from young star clusters into the nearby dense neutral gas.  (Although supernovae tend to inject more energy, they are thought to couple very poorly to dense, star-forming gas;  \citealt{matzner02}.)  Because emission line ratios reflect local conditions, this information would be complementary to the global analysis of \citet{at11}.    To this end we explored the properties of strongly radiation-confined layers using the analytical theory of Dr11 (Appendix \ref{Appendix:PlanarRegions}),  considered a host of physical effects and scenarios which could affect $\calU_{\rm obs}$ in real regions (\S~\ref{S:Dynamics} and \S~\ref{SS:Dust_abs}), and engaged in a suite of numerical models of dusty \HII regions in quasi-static force balance with an interior pressurized wind bubble (\S~\ref{S:model}).  Our models, though not meant to probe the full parameter space of \HII regions, are the first to explore a wide range of both the radiation force parameter ($\Gamma$ or $\Psi$) and the wind pressure parameter ($\Omega$).

Two effects complicate our plan to measure radiation pressure feedback using the apparent ionization parameter. 

\begin{figure}
\epsscale{1}
\plotone{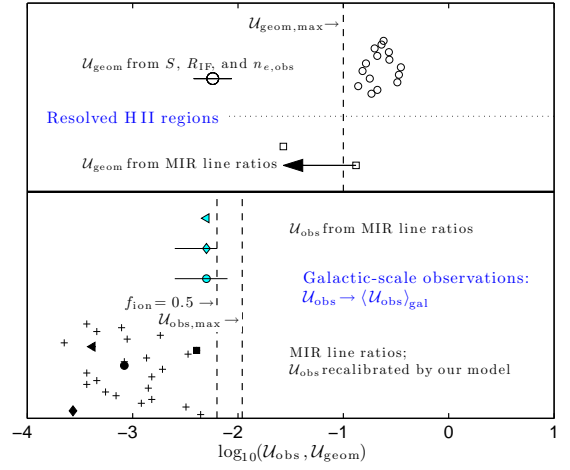}
\caption{
Samples of individual \HII regions and galactic-scale regions, plotted and divided into groups according to how $\calU_{\rm obs}$ is estimated. 
The thick horizontal line divides individual regions (upper portion) and galactic-scale regions (lower portion). 
We further distinguish the individual regions into two categories, based on whether $\calU$ is estimated  geometrically ($\calU_{\rm geom} = S/(4\pi R_{\rm IF}^2n_{i,{\rm obs}}c)$) or inferred from observed line ratios. 
Vertical dashed lines indicate theoretical maximum values $\calU_{\rm geom, max}$ (Appendix \ref{Appendix:PlanarRegions}, estimated from $\max(\Xi_{i, {\rm geom}})$) and $\calU_{\rm obs, max}$ (from [\NeIII]/[\NeII], figure \ref{fig:ne3ne2_2myr}), as well as the value of $\calU_{\rm obs}$ at which dust absorption becomes significant in the sense that $f_{\rm ion}=0.5$ in individual young regions with our fiducial parameters (Figure \ref{fig:ne3ne2_2myr}). 
Open circles -- single \HII regions in M82, \citet{mccrady07};
Large open circle -- \HII region M82-A1, \citet{smith06};
Open squares -- resolved \HII regions in the Antennae Galaxies, \citet{snijders07}; 
Cyan and black left-pointing triangle -- NGC 3256, \citet{thornley00};
Cyan and black diamond -- NGC 253, \citet{thornley00};
Cyan and black circle-- the central 500 pc starburst region of M82, where the bar indicates the possible range of $\log_{10} \calU_{\rm obs}$ modeled by \citet{fs03};
crosses -- recalibrated $\log_{10}\calU_{\rm obs}$ values of the galaxies in \citet{thornley00}.     \citet{snijders07} report a value which corresponds to our $\calU_{\rm geom}$, so we apply the approximate upper limit on this quantity.
}\label{fig:Uexamples}
\end{figure}

First is the fact that strongly radiation-dominated regions compress their ionized gas into thin layers whose internal pressure gradient balances the radiation force \citep{binette97}.   The ionization parameter recorded in any line ratio reflects the ratio of radiation to gas pressure where the lines are produced, and in radiation-confined regions $\calU_{\rm obs}$ saturates at a maximum value.  The resulting values of $\log_{10} \calU_{\rm obs, max}$ depend on the lines in question (\S~\ref{S:model}) and parameters such as the dustiness of the region and the ionizing spectrum, but are roughly $\sim -1.5$, the value for which $\Xi\simeq 1$.  For the particular neon line ratio plotted in Figure \ref{fig:ne3ne2_2myr}, the upper limit is actually $-1.97$.   Only significant levels of ram pressure or magnetic pressure, or serious departures from force balance, allow for $\calU_{\rm obs}$ to exceed this putative maximum (\S \ref{SS:Magnetic}-\ref{InertialConfinement}), and this point may be useful in diagnosing the physical states of unresolved regions.   Conversely any significant pressure on its inner boundary, such as the pressure due to shocked stellar winds, reduces $\Xi$ and $\calU_{\rm obs} $ (as well as the dust optical depth $\tau_d$) below their maximal values (\S~\ref{SS:Winds}).    Our limited compilation of ionization parameters in individual regions (figure \ref{fig:Uexamples}, top panels) is consistent with the proposition that $\calU_{\rm obs}$ varies up to the maximum imposed by radiation confinement; for instance, \citet{snijders07} find $\log_{10}\calU_{\rm obs} = -1.57$ in individual regions in the Antennae Galaxies. 

Another complicating effect is the suppression of the apparent ionization parameter on galactic scales by selective dust absorption of ionizing photons in the regions where $\calU$ is highest.  Gas in the local Universe is sufficiently dusty ($\gamma\simeq 10$) that the dust optical depth $\tau_d$ exceeds the radiation-to-gas pressure ratio $\Xi$, both of which vary proportionally to $\calU$: all radiation-confined regions (and some others) are dust-limited.  Such dust-limited regions convert a large portion of the stellar ultraviolet radiation into far infrared continuum, limiting the efficiency $f_{\rm ion}$ of line production.   For observations which span a wide range of local conditions, this biases the inferred estimate of $\calU$ (which we call $\calUgal$ in this case) toward regions of lower $\tau_d \leq 1$ and $\calU_{\rm obs}$.    

Within the toy model for galactic \HII region populations we present in \S \ref{SSS:Int_abs}, it appears this selection effect suppresses $\calUgal$ by a small amount ($<$0.05 dex) relative to an unbiased (i.e., starlight-weighted) average of $\calU_{\rm obs}$, primarily because the line efficiency $f_{\rm ion}$ gradually declines by only 0.5 dex (to a minimum of about 0.35) as radiation confinement becomes strong.   Although this result depends on the assumptions in our toy model, it strongly suggests that $\calUgal$ can be corrected for selective absorption and used to infer the physical characteristics of the galactic \HII region population.  Chief among these would be the ratio of radiation and gas-pressure forces averaged over a galaxy's ionized zones, a  quantity we plot (for the toy model) in the top panel of figure \ref{fig:UobsgalModel}.   

It has long been appreciated that the dust-bounded state contributes to the suppression of  fine-structure cooling lines relative to the far-infrared continuum in ULIRGs  \citep{1992ApJ...399..495V,1998PASP..110.1040B,abel09}, and that recombination luminosity under-represents the ionizing luminosity in this state; 

these effects provide observational checks on the degree of dust saturation one might infer from recombination line ratios. 

The galaxies collected in the bottom panel of figure \ref{fig:Uexamples} display an upper limit of $\log_{10}\calUgal\simeq -2.3$, with some uncertainty.  This is very close to the upper limit  of $-1.97$ we found for $\calU_{\rm obs}$ derived from a [\NeIII]/[\NeII] line ratio in young \HII regions (\S \ref{S:model}), especially considering the bias induced by dust attenuation.   We infer, therefore, that radiation pressure strongly modifies the ionized zones in these galaxies. 

A result of dust-induced bias is that the upper limit of $\calUgal$ will be weakly anti-correlated with metallicity. To test this, one must disentangle variations in $\calU$ from variations in the incident spectrum due to changes in stellar metallicity, and possible influences of metallicity on the stellar initial mass function (and more subtly, on stellar rotation and multiplicity).  

Moreover, our analysis of the radiation-confined state (Appendix \ref{Appendix:PlanarRegions}) suggests a  change in the properties of radiation-confined  \HII regions with metallicity above and below about $\log_{10}(Z/Z_\odot) = -1.2$,\footnote{Estimated assuming $T\propto Z^{-0.2}$, $\bar e_i\propto Z^{-0.04}$, and $\sigma_d\propto Z$.} corresponding to $\gamma=1$.  For higher values of $\gamma$, as we have already mentioned, dust absorption is significant in all radiation-confined zones.  Further, for $\gamma>1$ radiation-confined zones are compact (the density scale height is much smaller than the radius) and the Lyman-$\alpha$ line pressure is negligible.  None of these is true for $\gamma < 1$:  regions with significant radiation pressure extend inward toward the central source, are supported by line pressure as well as gas pressure, and are not significantly attenuated by dust.  Moreover, the wind pressure will typically be lower on account of the lower opacity of stellar atmospheres.   We leave a full investigation of this transition to future research, but point out that  many works on photo-ionization in active galactic nuclei have been carried out in the limit $\gamma=0$.  

We firmly rule out the hypothesis that resolved \HII regions are predominantly the ionized shells of energy-conserving wind bubbles (as envisioned for giant \HII regions by \citealt{1997MNRAS.289..570O}).  The pressure of shocked winds in such models lowers  $\calU$ well below observed values.   Although this discrepancy can be remedied by decimating the wind luminosity \citep{dopita05}, it is more natural to attribute the reduction in wind pressure to leakage of hot gas beyond the ionization front, either via direct outflow or radiative losses.   It is preferable to employ models which explicitly account for such leakage \citep[][KM09]{hc09}.   Such models form the basis for populations of \HII regions in the toy model of \S \ref{SSS:Int_abs} and in a future companion paper (Verdolini et al.\ in prep.), in which line emission by star-forming galaxies will be reconsidered in greater detail than is possible here.  

It is possible to interpret entire galaxies in terms of models of individual \HII regions, as we discussed in \S \ref{S:implication}.  This must be done with caution, because observations on galactic scales must necessarily average over a wide range of conditions, and are biassed by dust absorption.   Acknowledging this, one can nevertheless derive an effective upper limit on the wind pressurization $\Omega$ from the fact that a larger value of this parameter would limit $\calUgal$ below what is observed. 

For observational and theoretical studies of the ionization parameter, an important point is that the method used to infer $\calU_{\rm obs}$ affects the result, leading to systematic offsets between methods.  Traditionally line ratios are compared with suites of simulations in which $\calU_{\rm obs}$ represents the interior of an ionized layer (often one of uniform density), a definition which corresponds to our  $\calU_{\rm geom}$ so long as the layer is thin.   This definition encounters a couple difficulties.  First, the incident flux is attenuated by recombination and dust within the layer so that the characteristic value of $\calU$ is lower than the incident value (e.g., equation [\ref{UgeomInWeaver77}]).  Second, ionized gas can only be confined to a thin layer by an inner, pressurized region of hot gas, or (when $\gamma>1$) by strong radiation pressure; however, radiation confinement introduces a pressure and density gradient within the confined layer, and wind confinement limits the maximum values of $\Xi$ and $\calU$ well below unity and 0.1, respectively.  For these reasons, uniform-slab calculations are physically inconsistent at high values of $\calU$.  It is preferable to use suites of simulations involving spherical \HII regions in quasi-static force balance, but a single sequence of these excludes the role of a wind pressurization parameter (our $\Omega$).  

As an alternate standard, we advocate using, for $\calU_{\rm obs}$, the value of $\calU$ within a uniform mixture of gas photons for which the emitted line ratios match those observed.

\section{Acknowledgements}
We thank the anonymous referee for positive feedback, and C.F.~McKee and B.T.~Draine for insightful comments which helped to clarify the paper. 
SCCY thanks P.G.~Martin, G.J.~Ferland, K.~Blagrave, J.~Fischera, and P.~van Hoof for helpful discussions regarding photoionization codes.
CDM thanks A.~Bressan for an illuminating conversation.  This work is supported by a National Sciences and Engineering Research Council of Canada (NSERC) Grant, and a University of Toronto Fellowship.

\clearpage

\appendix

\section{A. Force balance and pressure equilibrium}  \label{Appendix:ForceBalance}
The momentum equation for fluid flow is 
\[ \ppt \rho \vect{v} + \divg \left(\rho \vect{vv} + P\vect{I} \right) =\frad + \fmag \] 
where $\vect{I}$ is the unit tensor and $\frad$ and $\fmag$ are the magnetic and radiation force per unit volume, respectively.  As elsewhere in the paper, $P$ (with no subscript) is the gas pressure.   Our analysis is based on a quasi-static idealization of an \HII region in which $v\rightarrow 0$ and $\ppt \rightarrow 0$, so this simplifies to a statement of force balance: 
\[ \nabla P = \frad + \fmag.\] 

As discussed by \citet{1984oup..book.....M}, $\frad = \divg\vect{P}_{\rm rad}$ where $\vect{P}_{\rm rad}$ is the radiation pressure tensor. In spherical symmetry, $\frad = -\ddr P_{\rm rad} -(3P_{\rm rad}-u_{\rm rad})/r$ for radiation pressure $P_{\rm rad}$ and energy density $u_{\rm rad}$.   Splitting the radiation into radially free-streaming and isotropic components $P_{\rm rad} = P_{\rm rad,fs}+ P_{\rm rad,is}$ with energy densities $u_{\rm rad} = P_{\rm rad,fs}+ 3P_{\rm rad,is}$, we have $\frad = -\ddr P_{\rm rad,is} - r^{-2} \ddr r^2 P_{\rm rad,fs}$.     Our free-streaming radiation is meant to represent attenuated radiation from the central source, whose luminosity at $r$ is $L(r)$ so that $P_{\rm rad, fs}(r) = L(r)/(4 \pi r^2 c)$, whereas the isotropic radiation field represents trapped line emission, especially Lyman $\alpha$, as well as the diffuse Lyman continuum radiation emitted during recombinations to the ground state. 

Inserting this and $\fmag = - \nabla P_{\rm mag} + (\vect{B}\cdot \nabla )\vect{B}/(4\pi)$ into the force balance equation  and writing only the radial component, 
\[ \ddr P = - \ddr P_{\rm rad,is} - r^{-2} \ddr r^2 P_{\rm rad,fs} - \ddr P_{\rm mag} + (\vect{B}\cdot\nabla)B_r. \] 
Writing $P_{\rm tot} = P + P_{\rm rad} + P_{\rm mag}$ and integrating across a range $r_{\rm in} \rightarrow r_{\rm in}+\Delta r $, 
\[ \Delta P_{\rm tot} = - 2 \int_{r_{\rm in}}^{r_{\rm in}+\Delta r} { L(r) \over 4\pi r^3 c} dr  
   + \int_{r_{\rm in}}^{r_{\rm in}+\Delta r}{(\vect{B}\cdot\nabla)B_r\over 4\pi}  dr. \] 
This shows that the change in $P_{\rm tot}$ is bounded across any range of $r$.   Since $L(r)$ is non-increasing, the magnitude of the first integral is maximized if radiation and matter do not interact, so that $L(r)$ remains constant: 
\[ 2 \int_{r_{\rm in}}^{r_{\rm in}+\Delta r}
            { L(r) \over 4\pi r^3 c} dr  
            \leq 2 P_{\rm rad, fs}(r_{\rm in})\left(1-{r_{\rm in}^2\over r_{\rm out}^2}\right)
             \longrightarrow 4 {\Delta r\over r_{\rm in} }P_{\rm rad, fs}(r_{\rm in})  \]
where the arrow represents the limit ${\Delta r\ll r_{\rm in} }$.   Moreover, the magnetic tension term $(\vect{B}\cdot\nabla)B_r/(4\pi)$ is of order $2P_{\rm mag}/R_c$ if $R_c$ is the curvature radius of the magnetic field.  When ionized gas is compressed in a thin shell of width $\Delta r$, $R_c$ likely to be intermediate between $\Delta r$ and $r$.  We can therefore say in general that when the shell is thin, 
\[ {|\Delta P_{\rm tot}| \over P_{\rm tot} }~~ {\buildrel{<}\over{\sim}}~~ 4 {\Delta r\over r} {P_{\rm rad, fs} {r_{\rm in}}\over P_{\rm tot}} 
+ 2 {\Delta r\over R_c} {P_{\rm mag}\over P_{\rm tot}},  \]
which is typically quite small: force balance implies near-constancy of the total pressure. 

\section{B. Dusty, radiation-dominated \HII regions: the planar limit}\label{Appendix:PlanarRegions}

We are interested in the properties of the Dr11 solutions in the limit that ionized gas is restricted to a geometrically thin shell, as occurs in radiation-dominated regions when $\GammaiS>1$ and (as we see below) $\gamma>1$.  We restrict equations (1)-(3) of Dr11 to the limit of a thin shell (constant $r$) and combine them to find
\begin{equation} 
{d\phi\over d\tau} = - \phi - \tilde{n}
\end{equation} 
(we use $\tau$ for the local dust optical depth, reserving $\tau_d$ for that of the entire region) and 
\begin{equation} 
{d\tilde{n} \over d\tau} = \gamma^{-1} \left( \phi + \beta e^{-\tau} + \tilde{n} \right)
\end{equation} 
where $\tilde{n} = n/n_{\rm ch}$ for a characteristic density 
\begin{equation} 
n_{\rm ch} = {   S \sigma_d \over 4 \pi R_{\rm IF}^2 \alpha }
\end{equation} 
which is defined so that, when $\tilde{n}=\phi=1$, dust absorption and ionization consume ionizing photons at equal rates. 

The solution with $\tilde{n}=0$ and $\phi=1$ at $\tau_d = 0$ is 
\begin{equation} \label{phi_planar} 
\phi(\tau) = { \gamma(\beta+1) e^{-\tau(\gamma-1)/\gamma} - \beta [(\gamma-1)e^{-\tau} + 1 ]-1 \over \gamma-1} 
\end{equation} 
and 
\begin{equation}\label{nhat_planar}
\tilde{n}(\tau) = {1+\beta \over \gamma-1}\left[1 - e^{-\tau(\gamma-1)/\gamma} \right], 
\end{equation} 
valid up to $\tau=\tau_{d}$, for which $\phi(\tau_d)=0$, so that  
\[ \gamma (\beta+1)e^{-k\tau_d} = \beta(\gamma-1)e^{-\tau_d} + \beta+1 \]
for $k=\gamma/(\gamma-1)$, which shows that when $\gamma \gg 1$ and $k\rightarrow 1$,  $\tau_d \rightarrow  \ln[(\gamma+\beta)/(\beta+1)]$.%
\footnote{A better approximation at moderate $\gamma$ is $\tau_d \simeq \ln(f) + 0.5e^{-f/8}$, where $f=(\gamma+\beta)/(\beta+1)$.}   Because of the planar geometry (Appendix \ref{Appendix:ForceBalance}) these solutions obey pressure equilibrium, $4\pi r^2 p = (L_n e^{-\tau} + L_i \phi)/c$, or in dimensionless terms,
\begin{equation} 
\gamma \tilde{n} + \phi + \beta e^{-\tau} = 1 + \beta
\end{equation} 
so that, at the IF, $\tilde{n} = [1 + (1-e^{-\tau_d})\beta]/\gamma$.   

In the special case $\gamma = 1$, we find 
\begin{equation} 
\phi(\tau) = (1+\beta)(1-\tau) - \beta e^{-\tau};  ~~~ \tilde{n}(\tau) = (1+\beta)\tau.
\end{equation} 

\begin{figure}
\epsscale{0.5}
\plotone{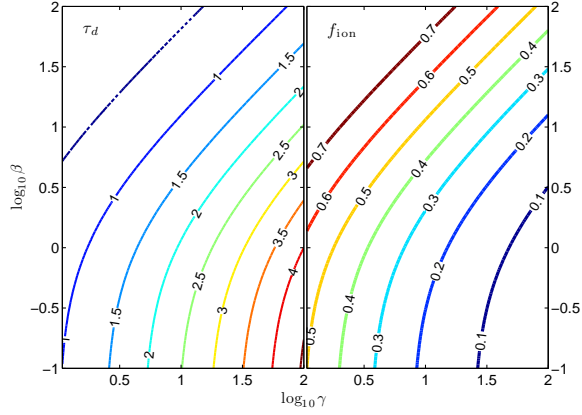}
\caption{Dust optical depth, emission-averaged ionizing radiation pressure-to-gas ratio, and fraction of ionizing radiation absorbed by gas within planar, radiation-confined, dusty \HII 
regions (equations \ref{phi_planar}, \ref{nhat_planar}, and  \ref{Xi_em_planar}) under the assumptions of Dr11: no pressure from recombination lines, constant temperature, identical dust cross sections for non-ionizing and ionizing radiation (whose luminosity ratio is $\beta$), and no pressurized inner region ($\Omega=0$).  The parameter $\gamma$ indicates the relative importance of dust in the consumption of ionizing photons in a state of equal gas and radiation pressures.  }\label{TaudXiFion}
\end{figure}

Within these solutions, 
\begin{equation}\label{Xi_em_planar}
\Xi_{i,{\rm em}} = {\int_0^{\tau_d} \phi \; d\tau \over \gamma \int_0^{\tau_d} \tilde{n}\; d\tau. }
\end{equation} 
and the fraction of ionizing starlight caught by gas is 
\begin{eqnarray}\label{fion_fdust_planar} 
f_{\rm ion} &=& 1-f_{\rm dust} = \int_0^{\tau_d} \tilde{n}\; d\tau \nonumber \\ 
&=& {1+\beta \over \gamma-1} \left[ {\tau_d } - {\gamma\over \gamma-1} \left(1-e^{ -{\gamma-1\over\gamma} \tau_d}\right)\right]\\
&\simeq& 1-\exp[0.7(\ln \tau_d - 1) - (\ln \tau_d)^2/6],
\end{eqnarray}
where the final approximation is correct to 2\% for the entire range of figure \ref{TaudXiFion}. 
In the very dusty limit $\gamma\gg1$, for which $\tau_d \sim 3$, the recombination luminosity is suppressed by a constant fraction which is roughly $(1+\beta)\tau_d/(\gamma-1)$. 

The radial coordinate can be determined by integrating $dr = d\tau/(n \sigma_d)$.  Because of the low density at the inside of the slab, radiation force is transmitted to the gas only through the dust, leading to a constant acceleration and (since the temperature is assumed to be constant) an exponential interior density distribution $n\propto \exp[ (1+\beta) n_{\rm ch} \sigma_d r/\gamma]$.  

Note, however, that equation \ref{nhat_planar} implies a negative density if $\gamma\leq 1$; this simply means  that the solution is incompatible with the imposed inner boundary of zero density and optical depth.  For such low dust opacity, the solution must extend inward to join the spherical solutions of Dr11, or meet an inner, pressurized region of hot gas.  The difference between compact ($\gamma>1$) and extended ($\gamma\leq 1$) radiation-confined regions  (with no inner hot gas) is apparent in the top panels of Dr11's figure 2.   For $\gamma\leq 1$, radiation-confined regions with $\GammaiS\gg1$ are still concentrated toward the ionization front, but the scaling of radiative acceleration with the neutral fraction allows gas to permeate the interior.   This effect would only be accentuated by the pressure of Lyman $\alpha$ line photons, which becomes significant when dust is scarce \citep{2004ApJ...608..282A}.

The net dust optical depth $\tau_d$ and ionization efficiency $f_{\rm ion}$, which reach their maximum and minimum values (respectively) in the radiation-confined state, are plotted in figure \ref{TaudXiFion}.    

\begin{figure}
\epsscale{0.5}
\plotone{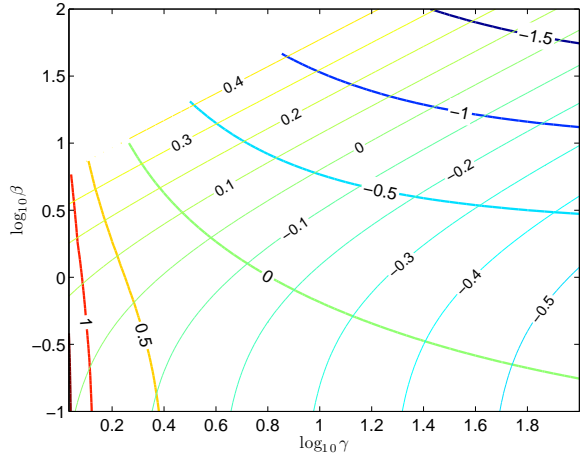}
\caption{Contours of $\log_{10} \Xi_{i, {\rm geom}}$ (thick lines) and $\log_{10} \Xi_{i,{\rm em}}$  (thin lines) in planar, radiation-confined ionized layers with no internal pressure.  Multiplied by $2.2 kT_i /\bar e_i$, these yield estimates for the upper limits of the geometrical and line-derived values of $\calU_{\rm obs}$, respectively.   However the latter is better estimated using the models of \S \ref{S:model} for the specific lines in question. }\label{XiEstimatesPlanar}
\end{figure}

In figure \ref{XiEstimatesPlanar} we plot two characteristic values of the ionizing radiation-to-gas ratio $\Xi_i$ in the radiation-confined state: the recombination-weighted average $\Xi_{i,{\rm em}}$, and the geometrical $\Xi_{i,{\rm geom}}$.  The latter is calculated using the emission-weighted density $n_{i, {\rm em}}$.    Estimates of $\calU_{\rm obs}$ and $\calU_{\rm geom}$ derived from these limits are only approximate, because line emission is not strictly proportional to the recombination rate. 

\end{document}